\documentstyle[12pt,epsf]{article}

\begin{document}

\sloppy

\begin{flushright}{UT-765\\ December '96}\end{flushright}

\vskip 1.5 truecm

\centerline{\large{\bf Large N expansion in global and local}}
\centerline{\large{\bf supersymmetric theories}}
\vskip .75 truecm
\centerline{\bf Tomohiro Matsuda
\footnote{matsuda@theory.kek.jp}}
\vskip .4 truecm
\centerline {\it Department of Physics, University of Tokyo}
\centerline {\it Bunkyo-ku, Tokyo 113,Japan}
\vskip 1. truecm

\makeatletter
\@addtoreset{equation}{section}
\def\theequation{\thesection.\arabic{equation}}
\makeatother

\vskip 1. truecm

\begin{abstract}
\hspace*{\parindent}
A systematic study of large N expansion in supersymmetric
theories are given.
Supersymmetric O(N) non-linear sigma model in two and three dimensions,
massless and massive supersymmetric QCD with $N_{f}<N_{c}-1$ and
supergravity models  are analyzed
in detail. 
Our main motivation  is to discuss how the previously studied
mechanism for dynamical generation of gaugino condensation
and superpotential is realized
in the framework of large N expansion.
\end{abstract}

\section{Introduction}
\hspace*{\parindent}
When one  extends the validity of the low energy effective field theory
to energy scales much higher than its characteristic mass scale,
 one faces 
to a scale hierarchy problem.
A typical example is the gauge hierarchy problem of the
Standard Model of the strong and electroweak interactions, seen as
a low-energy effective theory.
When the Standard Model is extrapolated to cut-off scales
$\Lambda\gg 1$Tev, there is no symmetry protecting the mass of
the elementary Higgs field from acquiring large value,
 and therefore the masses of the 
weak gauge bosons, receive large quantum corrections proportional
to $\Lambda$.
The most popular solution to the gauge hierarchy problem of the 
Standard Model is to extend it to a model with global
N=1  supersymmetry, effectively broken at a scale
$M_{Soft}\sim 1$Tev.
(See ref.\cite{Nills_rev} for a general review.)
These extensions of the Standard Model, for instance the 
Minimal Supersymmetric Standard Model(MSSM), can be
safely extrapolated up to cut-off scales much higher than the
electroweak scale, such as the supersymmetric unification
scale $M_{U}\sim 10^{16}$Gev, the string scale 
$M_{s}\sim 10^{17}$Gev,
or Planck scale $M_{P}=2.4\times 10^{18}$Gev.

To go beyond the MSSM, one must move to a more fundamental
theory with spontaneous supersymmetry breaking.
The  possible candidate for such a theory is N=1
supergravity coupled to gauge and matter fields, where 
the spontaneous breaking of local supersymmetry is not
incompatible with vanishing vacuum energy\cite{DB_rev}.
(Of course,  supersymmetry breaking can be
transmitted by gauge interaction\cite{Visible}.
But here we concentrate on the supergravity mediated supersymmetry
breaking models.)
In N=1 supergravity, the spin 2 graviton has as superpartner,
the spin 3/2 gravitino.
Here we consider the case that
the supersymmetry breaking is spontaneous, via the
super-Higgs mechanism\cite{Wess_Bagger}.
One is then bound to interpret the MSSM as an effective 
low-energy
theory derived from spontaneously broken supergravity.
The scale of soft supersymmetry breaking in MSSM, $M_{Soft}$,
is related to the gravitino mass $m_{\frac{3}{2}}$,
which sets the scale of the spontaneous breaking of local
supersymmetry.

The idea of breaking supersymmetry in a dynamical way was 
first presented
in ref.\cite{WittenIndex}.
In those articles a general topological argument was developed
in terms of the Witten index $Tr(-)^{F}$, showing that dynamical
supersymmetry breaking cannot be achieved unless there is chiral
matter or we include supergravity effects for which
the index argument cannot apply.
This is subsequently verified by explicitly
studying gaugino condensation in pure supersymmetric Yang-Mills,
a vector-like theory, for which gauginos condensate but do not
break global supersymmetry.
Breaking global supersymmetry with chiral matter was an open 
possibility in principle, but this approach ran into many
problems when tried to be realized in practice\cite{rep}.

The situation was improved very much with the coupling to 
supergravity.
The reason was that the simple gaugino condensation was found to be
sufficient to break supersymmetry once the coupling to gravity 
was included\cite{Nills_gaugino}.
This works in the hidden sector mechanism where gravity is the
messenger of supersymmetry breaking to the observable sector.
However, this mechanism does not work when the  gauge couplings 
are considered to be field dependent.
Non-perturbative effects, like gaugino condensations,
raises the moduli flat potentials, but
it is very difficult to obtain a phenomenological vacuum state.
The main difficulty lies in the runaway behavior of dilaton
potential\cite{NVK_Run,casas1}.
Gaugino condensation with field dependent gauge couplings
was anticipated and realized in a very natural way
in string theory.
The gauge 
couplings are functions of dilaton and moduli fields.
Furthermore, string theory provides a natural realization
of the hidden sector models by having a hidden sector\cite{GSW}.
Thus it is important to consider the following questions:
Does gaugino really condensate in supergravity theories?
If so, is runaway potential stabilized?
If runaway potential is not stabilized by gaugino condensation,
what effect should be responsible for the moduli fixing?
We know that an ordinary effective Lagrangian analysis 
cannot give us a satisfactory answer and we think   it is very important
to develop new ideas to analyze the dynamical properties
of supergravity theories.
(Assuming the scenario of two or more gaugino condensates,
the effective Lagrangian with confined hidden sector stabilizes 
the dilaton potential and breaks
supersymmetry with a more complicated dilaton superpotential
generated by multiple gaugino condensations\cite{casas1}.
However, one solution for the stabilization of the dilaton 
in the effective Lagrangian
requires a delicate cancellation between the contributions
from different gaugino condensates, which is not very natural.
The other solution generally requires the assistance of an 
additional
source of supersymmetry breaking\cite{DRSW,some7}.)
In this paper, we have developed a new method for the analysis
of the dynamical properties of supersymmetry theories.
Of course, as is recently discussed by many authors\cite{strong_string},
 we may find a solution to this problem by
introducing a new type of non-perturbative effects.
One of the very promising candidates is the effects of 
strongly coupled strings, but we do not consider
this possibility because at this moment we are not sure  how to handle and
apply this idea to phenomenological models.

In this paper, we analyze several types of supersymmetric models
by using  large N expansion.
In this limit, the relation between the effective Lagrangian
approach with confined hidden sector
and Nambu-Jona-Lasinio type approach is clarified.
In  large N limit, these two correspond to  different kinds 
of approximations of the exact solution.
Our main concern is to discuss how the previously mentioned 
mechanism for dynamical supergravity breaking is realized
in the framework of the large N expansion.
In Section 2, in order to discuss the applicability  of large N 
expansion in supersymmetric theories, we consider
a simple toy model.
This model, O(N) sigma model, is very useful when one introduces 
 large N expansion in supersymmetric theories.
In Section 3, we use this method to analyze the dynamical
properties of supersymmetric QCD.
Finally in Section 4, we study the dynamical supersymmetry
breaking in a local supersymmetric model.
The driving force for gaugino condensation, which is not
clear in other approaches, is now  obvious.
In general effective Lagrangian approach, the form of the potential
for  gaugino condensation is not well defined near 
the origin($\lambda\lambda\sim 0$) and it is difficult to understand
how this condensation takes place.
On the other hand, in Nambu-Jona-Lasinio type approach
or in the large N expansion, the potential is given by the wine-bottle
and we can easily imagine how fields rolls down to the condensating
vacuum. 

\newpage

\section{Supersymmetric O(N) non-linear sigma model}
\hspace*{\parindent}
People generally tend to think that, in supersymmetric theories,
no gap equations  should exist
because the non-renormalization theorem seems to 
ensure the cancellation of bosonic loops and fermionic ones.
This is true for the
superpotential motivated mass terms\cite{matsuda_gen}, 
but not applicable to
the D-term motivated ones, such as a gaugino soft mass.
Here, we should remember that N=1 four-dimensional
non-renormalization theorem says nothing about the renormalization
of D-terms but only about the superpotential renormalization
\cite{fujikawa}.

We believe it is very important to show, first of all,
that one can find a supersymmetric model in which we can exactly solve the 
gap-equation by means of large N expansion. 
Here we examine the phase structures of supersymmetric O(N)
non-linear sigma model in two and three dimensions.
We mainly follow refs.\cite{matsuda_on} and \cite{matsuda_soft}.
Some shortcomings of the previous papers on this topic
are corrected and deeper insights are given.

\subsection{Introduction}
\hspace*{\parindent}
Many years ago, Gross and Neveu\cite{gn}
have shown that dynamical symmetry breakdown 
is possible in asymptotically free field theories.
They obtain an expansion in powers of $1/N$ that is non-perturbative
in $g^{2}$.
This leads to a massive fermion and to a $\overline{\psi}{\psi}$
bound state at threshold.

Polyakov\cite{sigma}
 has pointed out that the $O(N)$ non-linear sigma model is 
asymptotically free and that the fundamental particle acquires a mass
for $N>2$.

Witten \cite{witten1}
has constructed a supersymmetric version of the two-dimensional
O(N) sigma model.
This is a hybridization of the non-linear sigma model and Gross-Neveu
model with Majorana fermions. 

Then one question appears naturally.
What is the difference between non-supersymmetric models and
supersymmetric ones?
If there is any difference, how is it realized?
Many authors tried to answer this question\cite{davis,alv}, but some 
questionable aspects of arguments are still left.

The purpose of this section is to clarify these ambiguities and present
a systematic treatment of this model.
To show explicitly what is going on, we are not going to eliminate
the auxiliary fields at the first stage by using  the 
equation of motion. 
If we eliminate all the auxiliary fields, it becomes difficult to find what 
relations we are dealing with.

\subsection{Review of the non-linear sigma model}
\hspace*{\parindent}

In this and the next subsections we are going to review the well-known
results of O(N) non-linear sigma model and four-fermion model to
fix the notations, and we show the strategy which is used 
throughout this paper.

The Lagrangian for O(N) sigma model is defined by
\begin{equation}
  L=-\frac{1}{2}n_{j}\partial^{2}n_{j}
\end{equation}
with the  local non-linear constraint
\begin{equation}
  n_{j}n_{j}=\frac{N}{g^{2}}.
\end{equation}
The sum over the flavor index j runs from 1 to N.
This constraint can be implemented by introducing a Lagrange 
multiplier $\lambda$.

Let us consider the Euclidean functional integral in the form:
\begin{eqnarray}
  Z&=&\int D\vec{n}\delta\left( (\vec{n})^{2}-\frac{N}{g^{2}} \right)
  exp\left( -\frac{1}{2}\int (\partial_{\mu}\vec{n})^{2}d^{D}x \right)
  \nonumber\\
  &=&\int D\lambda\int D\vec{n}exp\left\{ -\frac{1}{2}\int \left[
  (\partial_{\mu}\vec{n})^{2}+\lambda\left( (\vec{n})^{2}-
  \frac{N}{g^{2}} \right) \right] d^{D}x \right\}
\end{eqnarray}
The integral over $n$ is Gaussian and can be performed in a 
standard fashion. 
We have:
\begin{equation}
  Z=\int{D}\lambda{exp}\left(\frac{N}{2g^{2}}\int{\lambda}d^{D}x
  -\frac{N}{2}trln(-\partial^{2}+\lambda)\right)
\end{equation}
Let us first compute the variation of the action with respect to 
$\lambda$.
We get the following equation\cite{polyakov}:
\begin{eqnarray}
  \label{sigma_gap}
  \frac{N}{2g^{2}}&=&\frac{N}{2}\frac{\delta}{\delta\lambda(x)}trln
  (-\partial^{2}+\lambda(x))\nonumber\\
  &=&\frac{N}{2}G(x,x;\lambda)
\end{eqnarray}
Here we have introduced the Green function:
\begin{equation}
  G(x,x';\lambda)=<x'|(-\partial^{2}+\lambda)^{-1}|x>
\end{equation}
The meaning of the above equation becomes transparent if we
notice that
\begin{eqnarray}
  <n_{i}(x)n_{j}(y)>&=&Z^{-1}\int{D}\lambda\int{D}\vec{n}
  exp\left(-\frac{1}{2}\int\left\{(\partial\vec{n})^{2}+
  \lambda\left(\vec{n}^{2}-\frac{N}{g^{2}}\right)\right\}d^{D}x
  \right)\nonumber\\
  && \ \ \times{n}_{i}(x)n_{j}(y)\nonumber\\
  &=&\delta_{ij}\frac{\int{D}\lambda e^{-W}G(x,y;\lambda)}
  {\int{D}\lambda e^{-W}}\\
  W&=&-\frac{N}{2g^{2}}\int\lambda{d}^{D}x+\frac{N}{2}trln
  (-\partial^{2}+\lambda)\nonumber.
\end{eqnarray}
If $\lambda$ integration is to be approximated by the saddle point
$\lambda_{0}$, we obtain
\begin{equation}
   <n_{i}(x)n_{j}(y)>=\delta_{ij}G(x,y;\lambda_{0}).
\end{equation}
These equation shows that eq.(\ref{sigma_gap}) is nothing but the 
condition $<\vec{n}^{2}>=\frac{N}{g^{2}}$.
In other words, the gap equation can be obtained directly from the
constraint equation.
Here we call this simple calculation a tadpole method after
ref.\cite{tad}.
Now let us solve eq.(\ref{sigma_gap}).
Passing to the momentum representation, this ``gap equation''
is presented as:
\begin{eqnarray}
  \label{sigma_gapeq}
  G(x,x';\lambda_{0})&=&\int\frac{d^{D}p}{(2\pi)^{D}}\frac{e^{ip(x-x')}}
  {p^{2}+\lambda_{0}}\nonumber\\
  \frac{N}{g^{2}}&=&NG(x,x;\lambda_{0})\nonumber\\
  &=&N\int\frac{d^{D}p}{(2\pi)^{D}}\frac{1}{p^{2}+\lambda_{0}}.
\end{eqnarray}
This equation is applicable for any $D$ dimensions.
For D=2, we can obtain the precise form:
\begin{eqnarray}
  1 &=&\frac{g^{2}}{4\pi}log\frac{\Lambda^{2}}{\lambda_{0}}
  \nonumber\\
  \lambda_{0}&=&\Lambda^{2}exp\left(-\frac{4\pi}{g^{2}}\right)
\end{eqnarray}
For D=3, the situation differs  slightly.
We should include a critical coupling $g^{2}_{cr}$ defined by
\begin{equation}
  1=g^{2}_{cr}\int\frac{d^{3}p}{(2\pi)^{3}}\frac{1}{p^{2}}.
\end{equation}
If the coupling is strong($g^{2}>g^{2}_{cr}$), the gap equation has 
a non-trivial solution at $\lambda_{0}\ne0$.
(This critical coupling explicitly depends on the cut-off scale
$\Lambda$, so in three dimensions, we  regard this model
as a low-energy effective theory of some high-energy physics.
Of course one may find a good way to remove this cut-off dependence,
but here we do not consider such a detailed analysis.)
Using $g_{cr}$, we can rewrite (\ref{sigma_gapeq}) as:
\begin{eqnarray}
  \label{gap3}
  1&=&g^{2}\int\frac{d^{3}p}{(2\pi)^{3}p^{2}}-g^{2}\int
  \frac{d^{3}p}{(2\pi)^{3}}\left(\frac{1}{p^{2}}-\frac{1}{p^{2}
      +\lambda_{0}}\right)\nonumber\\
    &=&\frac{g^{2}}{g^{2}_{cr}}-g^{2}\int\frac{d^{3}p}{(2\pi)^{3}}
    \frac{\lambda_{0}}{p^{2}(p^{2}+\lambda_{0})}
\end{eqnarray}
The integral in (\ref{gap3}) is convergent.
Therefore, we have:
\begin{equation}
  m^{2}\equiv\lambda_{0}=const.\left(\frac{1}{g^{2}_{cr}}-
  \frac{1}{g^{2}}\right)^{2}
\end{equation}
If we take $g^{2}<g_{cr}^{2}$ something goes wrong with eq.(\ref{gap3}).
It does not have any solution, so the constraint 
$<\vec{n}^{2}>=\frac{N}{g^{2}}$
cannot be satisfied in this way.
To solve this puzzle,
we should also consider the possibility of spontaneous breaking 
of O(N) symmetry.
In above discussions, we have implicitly assumed that the 
vacuum expectation value of $\vec{n}$ would vanish.
Now we consider what will be changed if $\vec{n}$ itself gets a non-zero
vacuum expectation value.
Because of O(N) symmetry, the vacuum expectation value of
$\vec{n}\equiv(n_{1},n_{2},...n_{N})$ may be written as
\begin{equation}
  <\vec{n}>=(0,0,...\sqrt{N}v/g).
\end{equation}
So that the constraint equation (\ref{sigma_gapeq}) is changed.
\begin{eqnarray}
  \label{gapv}
  <(\vec{n})^{2}>&=&<\vec{n}>^{2}+<1-loop>\nonumber\\
  &=&N\left(\frac{v^{2}}{g^{2}}+\int\frac{d^{3}p}{(2\pi)^{3}}
  \frac{1}{p^{2}+\lambda_{0}}\right)=\frac{N}{g^{2}}
\end{eqnarray}
Of course, in two dimensions we cannot expect $\vec{n}$
to get any expectation value.
Considering the spontaneous breaking of $O(N)$ symmetry,
we introduce   another important 
critical coupling constant $g'_{cr}$.
\begin{equation}
  \frac{1-v^{2}}{g_{cr}^{'2}}=\int\frac{d^{3}p}{(2\pi)^{3}}
  \frac{1}{p^{2}}
\end{equation}
If $g$ is smaller than $g_{cr}$, then $v$ grows such that 
the constraint equation is satisfied in the weak coupling
 region($g'_{cr}\leq{g}\leq{g}_{cr}$)
 in a sense that not eq.(\ref{sigma_gapeq}) but eq.(\ref{gapv})
is satisfied by some $\lambda_{0}$. 
There appears, however, a peculiar ``flat direction''
in three dimensions.
As we have seen in the above analysis (\ref{gapv}), the constraint equation 
is satisfied for arbitrary values of $\lambda_{0}$ and $v_{0}$
once a certain relation is satisfied by these two variables.
To satisfy the constraint equation, $\lambda_{0}$ and $v$ 
should be related, but one cannot fix them at a unique point.
This relation defines the ``flat direction'' of such a peculiar
type.
This is because once the constraint equation is
satisfied, the potential term should vanish by definition.

As far as we deal with non-supersymmetric sigma model,
we have no primary reason to believe that the vacuum 
expectation value
of the field $v=<n_{j}>$ would vanish  in the strong coupling region.

\subsection{Review of the four-fermion model}
\hspace*{\parindent}
The four-fermion model is described by the Lagrangian
\begin{equation}
  \label{lag4}
  L=\frac{i}{2}\overline{\psi}_{j}\not{\! \partial}\psi_{j}
  +\frac{g^{2}}{8N}(\overline{\psi}_{j}\psi_{j})^{2}
\end{equation}
where the sum of the flavor index j runs from 1 to N and we
require that $g^{2}$ remain constant as N goes to infinity.
By introducing a scalar auxiliary field $\sigma$ 
($\sigma=\frac{N}{2g^{2}}\overline{\psi_{j}}\psi_{j}$) we may rewrite 
(\ref{lag4}) as
\begin{equation}
  L=\frac{i}{2}\overline{\psi}_{j}\not{\! \partial}\psi_{j}
  +\frac{1}{2}\sigma\overline{\psi}_{j}\psi_{j}-\frac{N\sigma^{2}}
  {2g^{2}}.
\end{equation}
Let us consider the functional integral in the form:
\begin{equation}
  Z=\int{D}\psi_{j}D\sigma{exp}\left[\int{d}^{D}x\left\{
  \frac{1}{2}\overline{\psi}_{j}(i\not{\! \partial}+\sigma)\psi_{j}
  -\frac{N}{2g^{2}}\sigma^{2}\right\}\right]
\end{equation}
Integrating over the field $\psi_{j}$ we get an effective action
for the field $\sigma$:
\begin{equation}
  Z=\int{D}\sigma{exp}\left[-\frac{N}{2g^{2}}\int{d}^{D}x
  \sigma^{2}+\frac{N}{2}Trln(i\not{\! \partial}+\sigma)\right]
\end{equation}
We impose the stationary condition which gives the gap equation.
\begin{equation}
  \frac{N<\sigma>}{g^{2}}-\frac{N}{2}\int\frac{d^{D}p}{(2\pi)^{D}}
  tr\frac{1}{-\not{\! p}+<\sigma>}=0
\end{equation}
As is in non-linear sigma model discussed in the previous part of this
section,
this gap equation represents the condition
\begin{equation}
  \label{diff}
  \frac{N}{g^{2}}<\sigma>=
    \frac{1}{2}<\overline{\psi}_{j}\psi_{j}>|_{m_{\psi}=<\sigma>}.
\end{equation}
where non-zero $\sigma$ corresponds to the condensation of 
$\overline{\psi}_{j}\psi_{j}$.
For D=2, we obtain the equation:
\begin{eqnarray}
  \frac{1}{g^{2}}&=&\int\frac{d^{2}p}{(2\pi)^{2}}
  \frac{1}{p^{2}+<\sigma>^{2}}\nonumber\\
  <\sigma>^{2}&=&\Lambda^{2}exp\left(-\frac{4\pi}{g^{2}}\right)
\end{eqnarray}
For D=3, we have a critical coupling constant.
The saddle point exists only within the branch
\begin{equation}
  0<\frac{1}{g^{2}}\leq\frac{1}{g_{cr}^{2}}
\end{equation}
where
\begin{equation}
  \frac{1}{g^{2}_{cr}}
  \equiv\int\frac{d^{3}p}{(2\pi)^{3}}\frac{1}{p^{2}}.
\end{equation}
The crucial difference from the $O(N)$ non-linear sigma model is
that eq.(\ref{diff}) always has a trivial solution at $\sigma=0$.
In the weak coupling region, non-trivial saddle point
vanishes but the trivial solution always exists.

\subsection{Phases in Supersymmetric Non-Linear Sigma Model}
\hspace*{\parindent}
Supersymmetric non-linear sigma model is usually defined by
the Lagrangian
\begin{equation}
  L=\frac{1}{2}\int{d}^{2}\theta(\overline{D}\Phi_{j})(D\Phi_{j})
\end{equation}
with the non-linear constraint
\begin{equation}
  \label{const2}
  \Phi_{j}\Phi_{j}=\frac{N}{g^{2}}.
\end{equation}
where the sum of the flavor index j runs from 1 to N.
The superfields $\Phi_{j}$ may be expanded  in components
\begin{equation}
  \Phi_{j}=n_{j}+\overline{\theta}\psi_{j}+\frac{1}{2}
  \overline{\theta}\theta{F}_{j}
\end{equation}
and we define the super covariant derivative of the form: 
\begin{equation}
  D=\frac{\partial}{\partial\theta}-i\overline{\theta}\not{\! \partial}
\end{equation}
In order to express the constraint (\ref{const2}) as a $\delta$
function, we introduce a Lagrange multiplier superfield $\Sigma$.
\begin{equation}
  \Sigma=\sigma+\overline{\theta}\xi+\frac{1}{2}\overline{\theta}
  \theta\lambda
\end{equation}
We thus arrive at the manifestly supersymmetric action for the 
supersymmetric sigma model.
\begin{equation}
  \label{lag3}
  S=\int{d}^{D}xd^{2}\theta\left[\frac{1}{2}(\overline{D}\Phi_{j})(D\Phi_{j})
  +\frac{1}{2}\Sigma\left(\Phi_{j}\Phi_{j}-\frac{N}{g^{2}}
  \right)\right]
\end{equation}
where D=2,3.
In component form, the Lagrangian from (\ref{lag3}) is
explicitly written in the following form. 
\begin{eqnarray}
  L&=&-\frac{1}{2}n_{j}\partial^{2}n_{j}+\frac{i}{2}
  \overline{\psi}_{j}\not{\! \partial}\psi_{j}+\frac{1}{2}
  F_{j}^{2}
  -\sigma{n}_{j}F_{j}-\frac{1}{2}\lambda{n}_{j}^{2}\nonumber\\
   &&+\frac{1}{2}\sigma\overline{\psi}_{j}\psi_{j}+\overline{\xi}
  \psi_{j}n_{j}+\frac{N}{2g^{2}}\lambda
\end{eqnarray}
We can see that $\lambda, \xi,$ and $\sigma$ are the respective
Lagrange multipliers for the constraints:
\begin{eqnarray}
  \label{what}
  n_{j}n_{j}&=&\frac{N}{g^{2}}\nonumber\\
  n_{j}\psi_{j}&=&0\nonumber\\
  n_{j}F_{j}&=&\frac{1}{2}\overline{\psi}_{j}\psi_{j}
\end{eqnarray}
The second and the third constraints are obtained by the supersymmetric 
transformations of the first one.
We must not include kinetic terms for the field $\sigma$ and $\xi$
so as to keep these constraints manifest.
We can examine these constraints in a manner we have used 
in the previous part of this section.
As we have seen, we can directly solve the gap equation without calculating
the explicit 1-loop potential.
Below, we analyze each constraint and solve gap equations.

First we examine the two dimensional theory.

(1) Scalar part\\
\begin{equation}
  \label{1scalar}
  <n_{j}n_{j}>|_{m_{n}^{2}=<\lambda>+<\sigma^{2}>}=\frac{N}{g^{2}}
\end{equation}

In two dimensions, as we have seen above, $m_{n}$ is always 
non-zero.
\begin{eqnarray}
  m_{n}^{2}&=&<\lambda>+<\sigma>^{2}\nonumber\\
  &=&\Lambda^{2}exp\left(-\frac{4\pi}{g^{2}}\right)
\end{eqnarray}
This fixes $m_{n}$ at a dynamical scale but does not fix
$<\lambda>$ and $<\sigma>^{2}$ independently.

(2) Fermion part\\
\begin{equation}
  \label{2fermi}
  <n_{j}F_{j}>=\frac{1}{2}<\overline{\psi}_{j}{\psi}_{j}>
\end{equation}
One may think that the fermionic condensation
should vanish to keep supersymmetry unbroken, but this notion is
not always true.
This relation includes auxiliary field $F_{j}$, to be
eliminated by equation of motion.
After substituting $F_{j}$ by $\sigma{n}_{j}$, we obtain
at one-loop level: 
\begin{eqnarray}
  <n_{j}F_{j}>&=&<\sigma n_{j}n_{j}>\nonumber\\
  &=&<\sigma><n_{j}n_{j}>=\frac{1}{2}<\overline{\psi}_{j}\psi_{j}>
\end{eqnarray}
If we impose the O(N) symmetric constraint 
$<n^{2}>=\frac{N}{g^{2}}$, we have
\begin{eqnarray}
  \label{22}
  \frac{N}{g^{2}}<\sigma>&=&\frac{1}{2}<\overline{\psi}\psi>|_{
    m_{\psi}=<\sigma>}\nonumber\\
  \frac{N}{g^{2}}&=&\frac{N}{2}\int\frac{dp^{D}}{(2\pi)^{D}}
  \frac{1}{p^{2}+<\sigma>^{2}}.
\end{eqnarray}
For D=2, the solution is
\begin{equation}
  \label{2kai}
  <\sigma>^{2}=\Lambda^{2}exp\left(-\frac{4\pi}{g^{2}}\right).
\end{equation}
Substituting $<\sigma>$ in the first constraint (\ref{1scalar}) with 
(\ref{2kai}), we can find that $<\lambda>$ 
must vanish(in this point ref.\cite{alv} was wrong).
This means that the field $\psi$ gains the same mass as $n$,
and simultaneously supersymmetric order parameter $<\lambda>$ 
vanishes.
We can say that the supersymmetry is not broken in two dimensions
as is predicted by Witten\cite{WittenIndex}.
Moreover, we can examine the assumption of vanishing $v$ by 
decomposing the constraint(\ref{2fermi}) as follows.
\begin{equation}
  <\sigma>\left(v^{2}+\int\frac{d^{2}p}{(2\pi)^{2}}\frac{1}{p^{2}
    +<\sigma>^{2}}\right)
  =\int\frac{d^{2}p}{(2\pi)^{2}}\frac{<\sigma>}
  {p^{2}+<\sigma>^{2}}
\end{equation}
Bosonic  and fermionic loops exactly cancel. Finally we obtain:
\begin{equation}
  \label{sv}
  <\sigma>v^{2}=0
\end{equation}
As $<\sigma>$ is non-zero in two dimensions(\ref{2kai}),
 we must set $v=0$.

When D=3, we can find a solution for the eq.(\ref{1scalar}) only in the 
region  $g>g'_{cr}$.
The critical coupling  $g'_{cr}$ is defined by
\begin{equation}
  \frac{1-v^{2}}{g_{cr}^{'2}}=\int\frac{d^{3}p}{(2\pi)^{3}}
  \frac{1}{p^{2}}
\end{equation}
while $g_{cr}$ is defined as:
\begin{equation}
   \frac{1}{g^{2}_{cr}}
  \equiv\int\frac{d^{3}p}{(2\pi)^{3}}\frac{1}{p^{2}}.
\end{equation}
O(N) symmetry is expected to be spontaneously broken 
 in the region $g'_{cr}<g<g_{cr}$ by a non-zero value of $v$. 
And when $g=g'_{cr}$, $m_{n}$ would vanish.

For the fermionic part(\ref{2fermi}), in D=3,
we also have a critical coupling constant.
As far as $g\geq{g}_{cr}$, we have nothing to worry about.
In this strong coupling region, both supersymmetry and 
O(N) symmetry are preserved in a fashion like two dimensions.
In this region $v$ cannot develop any non-zero value
because eq.(\ref{sv}) is also true for the strongly coupled
three dimensional theory.  
However, in the weak coupling region, something goes wrong.
There is no non-trivial solution for fermionic constraint(\ref{2fermi}) and 
there is no fermionic condensation (This means that the only
possible solution is $<\sigma>=0$).
Thus we can see from eq.(\ref{sv}) that $v$ can develop non-zero
value in this weak coupling region.
This is supported by the constraint (\ref{1scalar}), because this 
does not have any solution in the weak coupling region
unless we allow $v$ to develop non-vanishing value. 
Eq.(\ref{gapv}) suggests:
\begin{equation}
  v^{2}=1-g^{2}\int\frac{d^{3}p}{(2\pi)^{2}}\frac{1}{p^{2}}
\end{equation}
Naive consideration also supports this analysis.
In general, we can expect that quantum effects in 
correlation functions like 
$<n_{j}n_{j}>$ or $<\overline{\psi}\psi>$
would vanish in the weak coupling limit. 
But we have a O(N) symmetric constraint.
It is natural to think that the field $n$ itself gains 
 expectation value to complement quantum effects.
This simply means  that  classical effects become more dominant
in the weak coupling region, therefore the O(N) symmetric
constraint is satisfied classically.
(i.e. in the weak coupling limit $g\rightarrow0$ we obtain $v=1$.
This is a classical solution of the constraint.)
As a result, in the weak coupling region O(N) symmetry is
spontaneously broken by non-zero value of $v$.

We should also note that, in the weak coupling region, there is 
also a possible solution of non-zero $\lambda_{0}$ 
if $g\ne g'_{cr}$. 
It induces a supersymmetry breaking term to the Lagrangian of the form:
\begin{equation}
  L_{break}=\lambda_{0}\left((\vec{n})^{2}-\frac{N}{g^{2}}
  \right)
\end{equation}
On the constrained phase 
space($(\vec{n})^{2}-\frac{N}{g^{2}}=0$), vacuum energy also 
seems to vanish for non-zero $\lambda_{0}$
as far as $v$ valances to satisfy the constraints.
(See section 2.2.)
Can we think that there remains some unusual type of
flat direction, with non-zero value of F-component?
Of course this statement is wrong.
After including effective kinetic terms,
$\sim\lambda\lambda$ appears effectively(see ref.\cite{maha}).
Then,  we can find positive vacuum energy for 
supersymmetry breaking phase($\lambda\ne0$)
as in the usual type of supersymmetric theories.
So we can conclude: 

(1) In two dimensions, both supersymmetry and O(N) symmetry 
are not broken. 
This means that  $\lambda$  and $v$ remain zero for any
value of $g$.

(2) In three dimensions, both supersymmetry and O(N) symmetry 
are not broken (i.e. $\lambda$  and $v$ remain zero) in the strong 
coupling region.
O(N) symmetry can be broken in the weak coupling region, but
supersymmetry is kept unbroken in both phases.

\subsection{Some peculiar properties of the present model}
\hspace*{\parindent}
In this section we discuss the stability of the
dynamically generated supersymmetric mass terms against
the supersymmetry breaking mass term.

We examine the supersymmetric non-linear O(N) sigma model
with a supersymmetry breaking mass term.
In two dimensions, we will find that
the mass difference between supersymmetric partner fields vanishes
by the quantum effect. 
In three dimensions, the mass difference vanishes 
in the strong coupling region,
 but O(N) symmetry is always broken both in strong and weak coupling
region.

To see what happens, now let us extend the above analysis to
include a supersymmetry breaking mass term.
Here we consider a supersymmetry breaking mass term of the form:
\begin{equation}
  \label{soft}
  L_{break}=\frac{-1}{2}m_{s}^{2} n_{j}^{2}
\end{equation}
Including this supersymmetry breaking mass term, we can  
calculate the gap equation explicitly.
For the scalar part it becomes:
\begin{eqnarray}
  n_{j}n_{j}|_{m_{n}^{2}=<\lambda>+<\sigma>^{2}+m_{s}^{2}}&=&
  N\int\frac{d^{D}p}{(2\pi)^{D}}
  \frac{1}{p^{2}+<\lambda>+<\sigma>^{2}+m_{s}^{2}}
  \nonumber\\
  &=&\frac{N}{g^{2}}
\end{eqnarray}
The fermionic part is unchanged by the supersymmetry  breaking mass term.
For $D=2$ we can solve this equation explicitly.
\begin{eqnarray}
  \label{soft_gap2}
  1&=&\frac{g^{2}}{4\pi}log\frac{\Lambda^{2}}
  {<\lambda>+<\sigma>^{2}+m_{s}^{2}}
  \nonumber\\
  m_{n}^{2}&=&
  <\lambda>+<\sigma>^{2}+m_{s}^{2}\nonumber\\
  &=&\Lambda^{2}exp\left(-\frac{4\pi}{g^{2}}\right)
\end{eqnarray}
$<\sigma>^{2}$ is determined by the fermionic part which is unchanged by 
the supersymmetry breaking term(\ref{soft}).
\begin{eqnarray}
  m_{\psi}^{2}&=&
  <\sigma>^{2}\nonumber\\
  &=&\Lambda^{2}exp\left(-\frac{4\pi}{g^{2}}\right).
\end{eqnarray}

These two equations suggest two consequences.
One is that the order parameter for supersymmetry breaking($\lambda$)
 gets non-zero value.
The shift of $\lambda$ that is induced by the supersymmetry breaking
 mass term is:
\begin{equation}
  <\lambda>+m^{2}_{s}=0
\end{equation}
And  supersymmetry is broken.
The second consequence  is more peculiar.
As we can see from explicit calculations,
 dynamically generated masses are unchanged and the mass degeneracy
is not removed by the explicit supersymmetry breaking  mass term.
This happens because the auxiliary field $\lambda$ has absorbed 
$m_{s}$ so that the two masses balance.

We conclude that, if we believe the validity of the large N expansion, the
dynamical masses are unchanged while the supersymmetry breaking parameter
develops non-zero value.

The crucial point of our observation lies in the fact that
we can absorb the supersymmetry breaking mass term by redefining a field.
The simplest and trivial example is the ordinary O(N) non-linear
sigma model with an explicit mass term.
This is written as:
\begin{equation}
  L=-\frac{1}{2}n_{j}\partial^{2}n_{j}
  -\frac{1}{2}\lambda(n_{j}^{2}-\frac{N}{g^{2}})
  -m^{2}_{s}n_{j}^{2}
\end{equation}
Does the explicit mass term changes the dynamical mass?
The answer is no.
This can easily be verified by redefining $\lambda$ as
$\lambda'=\lambda+m^{2}_{s}$.
Lagrangian is now:
\begin{equation}
  L=-\frac{1}{2}n_{j}\partial^{2}n_{j}-
  \frac{1}{2}\lambda'(n_{j}^{2}-\frac{N}{g^{2}})
  -\frac{N}{2g^{2}}m^{2}_{s}
\end{equation}
We can find that the mass term is absorbed in $\lambda'$ and
only a constant is left.
Of course, this constant does not change the gap equation.

In three dimensions, however, it is not so simple.
Many fields and their equations form complex relations and determine 
their values each other.
Let us see more details.
In three dimensions, things are not so easy.
As is discussed above, 
this model has a weak coupling region where no dynamical mass is produced
so  no balancing effect between superpartner masses
works in this region.
Setting $\lambda=0$, we find $m_{n}=m_{s}$ and $m_{\psi}=0$ when $g$ is
small.
This agrees with the naive expectation.
What will happen if we go into the strong coupling region
where the gap equations develop non-trivial solutions and
  the fermion becomes massive?
If there were no supersymmetry breaking mass term, O(N) symmetry 
restoration occurs in this region.
But because $m_{s}$ is non-zero, $v$ must develop non-zero value
in order to compensate $m_{s}$ and 
satisfy the constraint equation(\ref{gapv}).
In this case, we cannot set $\lambda=0$ because $\lambda$ and $v$
should be determined by minimizing  the full 1-loop potential.

To summarize, after adding a breaking term, some
fields slide to compensate $m_{s}$ but the mechanism is not trivial.
Even in our simplest model, many complex relations determine their
values.
In two dimensions, we found that
the mass difference between supersymmetric partner fields vanishes
 but the supersymmetry is still broken in a sense that $<\lambda>$
is non-zero. 
In three dimensions, the mass difference is always non-zero.
In the weak coupling region it is $m_{s}$,
but to determine the precise value of the mass difference
 in the strong coupling region,
 we should solve the following equation:

\begin{eqnarray}
  \label{soft3}
  n_{j}n_{j}|_{m_{n}^{2}=<\lambda>+<\sigma>^{2}+m_{s}^{2}}
  &=&v^{2}+N\int\frac{d^{3}p}{(2\pi)^{3}}
  \frac{1}{p^{2}+<\lambda>+<\sigma>^{2}+m_{s}^{2}}
  \nonumber\\
  &=&\frac{N}{g^{2}}
\end{eqnarray}
where $\sigma$ should be determined by the fermionic constraint.
This equation (\ref{soft3}) determines the relation between $v$
and $\lambda$.
To go further, we should minimize the full 1-loop potential
for $v$ and $\lambda$ under the constraint(\ref{soft3}).
This calculation is too complicated to find exact solutions,
but it is obvious that $v$ and $\lambda$ are simultaneously non-zero.
So we can conclude that in the strong coupling region, mass difference
is non-zero and O(N) symmetry is also broken. 
In three dimensions  O(N) symmetry is always broken
and mass difference is always non-zero. 
Its value (mass difference) coincides with $m_{s}$ in the
weak coupling region, but in the strong coupling region it
is determined by very complicated relations. 

What is new in this section  is:

1) It was previously believed that the fermionic condensation
is parameterized by  $\lambda$ which is the F-component
of the Lagrange multiplier.
In ref.\cite{alv}, it is discussed that even if fermionic condensation
occurs and $\lambda$ becomes non-zero, supersymmetry is still
maintained since the dynamical masses for fermion and scalar
field are equal.
By using tadpole method, we showed that $\lambda$ should vanish
in this theory.

2) If one calculates a naive 1-loop effective potential, one
will find a fictitious negative energy solution at $\lambda\ne 0$.
This problem can be evaded if  effective $\lambda\lambda$ term
is included.

3) We analyzed  the effects of the  supersymmetry breaking mass term
on dynamical properties.

Recently, many groups have discussed the dynamical properties of softly
broken supersymmetric theories\cite{rec}, but some non-trivial
assumptions were needed.
(For example, they  assumed that the non-perturbative 
superpotential $W^{np}$ is not  changed by the small soft mass.
This is a very strong assumption which should be verified
in another way. 
To be more specific, in  O(N) non-linear sigma model this assumption
corresponds to:
Neglect the $m_{s}$ factor in the gap equations (\ref{soft_gap2}) and
(\ref{soft3}).
Of course, this is obviously  wrong.)
Although our model is much simpler, we think it is important to 
consider the  explicitly solvable models as a toy model for these
complicated issues.

\subsection{Summary of section 2}
\hspace*{\parindent}
In this section we have studied supersymmetric O(N) non-linear
sigma model in two and three dimensions.
Supersymmetric O(N) non-linear sigma model is already analyzed
in refs.\cite{alv,maha} by using 1-loop effective potential, but some
uncertainties were left. (See 1),2) and 3) in the last part of the
previous subsection.)
Using tadpole method, we have re-analyzed this theory
and found that this method is very useful especially
when we analyze large N expansions in  supersymmetric theories.
Many uncertainties are resolved.
We have also considered the effect of the supersymmetry breaking mass term
and its peculiar properties.

\newpage
\section{Dynamical analysis in supersymmetric gauge theories}
\hspace*{\parindent}

\subsection{General Analysis}

\subsubsection{Supersymmetric pure Yang-Mills theory}
\hspace*{\parindent}

Let us first review the dynamical analysis of supersymmetric Yang-Mills
theory.
In this paper, the word pure supersymmetric Yang-Mills theory(SYM)
means the supersymmetric gauge theory with $SU(N_{c})$ gauge group
without matter field.

The Lagrangian is generally  written as:
\begin{eqnarray}
  L=\frac{1}{g^{2}}\int d^{2}\theta W^{\alpha a}W^{\alpha}_{a}+h.c.
\end{eqnarray}
where the chiral representation of the vector superfield ($W^{\alpha}$)
 contains
gauge boson($A_{\mu}$) and its fermionic partner($\lambda^{\alpha}$).
\begin{eqnarray}
  W_{\alpha}^{a}&=&-i\lambda_{\alpha}^{a}+\left[\delta_{\alpha}^{\beta}
  D^{a}-
  \frac{i}{2}(\sigma^{\mu}\overline{\sigma}^{\nu})_{\alpha}^{\beta}
  (\partial_{\mu}A^{a}_{\nu}-\partial_{\nu}A^{a}_{\mu}+
   f^{abc}A^{b}_{\mu}A^{c}_{\nu})
\right]\theta_{\beta}
  \nonumber\\
  &&+\theta\theta(\sigma^{\mu})_{\alpha\dot{\alpha}}\partial_{\mu}
  \overline{\lambda}^{a\dot{\alpha}}
\end{eqnarray}
It is known that Witten index is non-zero in this theory,
 so we can expect
that supersymmetry should not be broken.
On the other hand, it is also known that gaugino
should condensate and chiral symmetry is broken.
This  condensation can be checked  by the instanton calculation in the 
weak coupling region or by using the effective (confined) Lagrangian
analysis in the strong coupling region.
These analyses present consistent results so we can believe that
gaugino condensation  occurs in supersymmetric pure
Yang-Mills theory at any scale.

For later convenience, here we explicitly calculate the 
effective potential for the composite field $U=\frac{g^{2}}{32\pi^{2}}
W^{\alpha}W_{\alpha}$.
Other approaches, such as the instanton calculation,
are entirely reviewed in ref.\cite{rep}.
The underlying principle of this construction is t'Hooft's anomaly
matching condition, which demands that the effective low energy Lagrangian,
valid below some scale $\Lambda$, should reproduce the anomalies of the
underlying constituent theory.
In the case of the pure supersymmetric Yang-Mills model it is well known that
R-symmetry and supersymmetry current as well as the energy momentum tensor
lie in the same supersymmetry multiplet.

In terms of constituent fields the lowest component of $U$ is proportional
to the gaugino bilinear $\lambda\lambda$, so it makes sense to take
$U$ as the goldstone multiplet entering the low energy Lagrangian.
The R-symmetry acts as
\begin{equation}
  U(x,\theta)\rightarrow e^{3i\alpha}U(x,e^{-3i\alpha}\theta)
\end{equation}
while the scale symmetry acts as 
\begin{equation}
  U(x,\theta)\rightarrow e^{3\gamma}U(xe^{\gamma},\theta e^{\gamma/2})
\end{equation}
Assuming that the anomalies associated with the above classical 
invariances are reproduced by the superpotential $W$, one obtains
the holomorphic equation for the superpotential (note that the
 imaginary part of $U$ contains $\tilde{F}F$)
\begin{itemize}
\item Anomaly from fundamental Lagrangian
  \begin{eqnarray}
    \delta L&=&2N_{c}\alpha \frac{g^{2}}{32\pi^{2}}F\tilde{F}
    \nonumber\\
    &=&2N_{c}\alpha i\int d^{2}\theta U+h.c.
  \end{eqnarray}
\item Anomaly from composite Lagrangian
  \begin{equation}
    \int d^{2}\theta(-2i\alpha W + 2i\alpha U\frac{dW}{dU})+h.c.
  \end{equation}
\end{itemize}
The resulting anomaly matching condition is:
\begin{equation}
  U\frac{\partial W}{\partial U}-W=N_{c}U,
\end{equation}
which has the general solution of the form
\begin{equation}
  W=aU+N_{c}Ulog\left(\frac{U}{\mu^{3}}\right)
\end{equation}
where $a$ is in general undetermined and corresponds to the rescaling
of the condensation scale $\mu$, and b is a function of the
gauge coupling which is fixed by the anomaly constraints.

To examine this model explicitly, we need some information on the
K\"ahler potential $K$.
Usually one demands that the variation of the D-component of K is 
non-anomalous and it fixes its form to be:
\begin{equation}
  K=const.(U\overline{U})^{1/3}
\end{equation}
(If one takes $K=K(U\overline{U})$ then the variation vanishes
identically for R-symmetry.
However the variation under scale symmetry suggests that
\begin{equation}
 \delta K|_{D}=\gamma \left[3\frac{\partial K}{\partial U}U+
 3\frac{\partial K}{\partial \overline{U}}\overline{U}-2K\right]_{D}
\end{equation}
where only  $K=const.(U\overline{U})^{1/3}$ is allowed.)

\begin{figure}[hp]
  \epsfxsize=13cm

  \centerline{\epsfbox{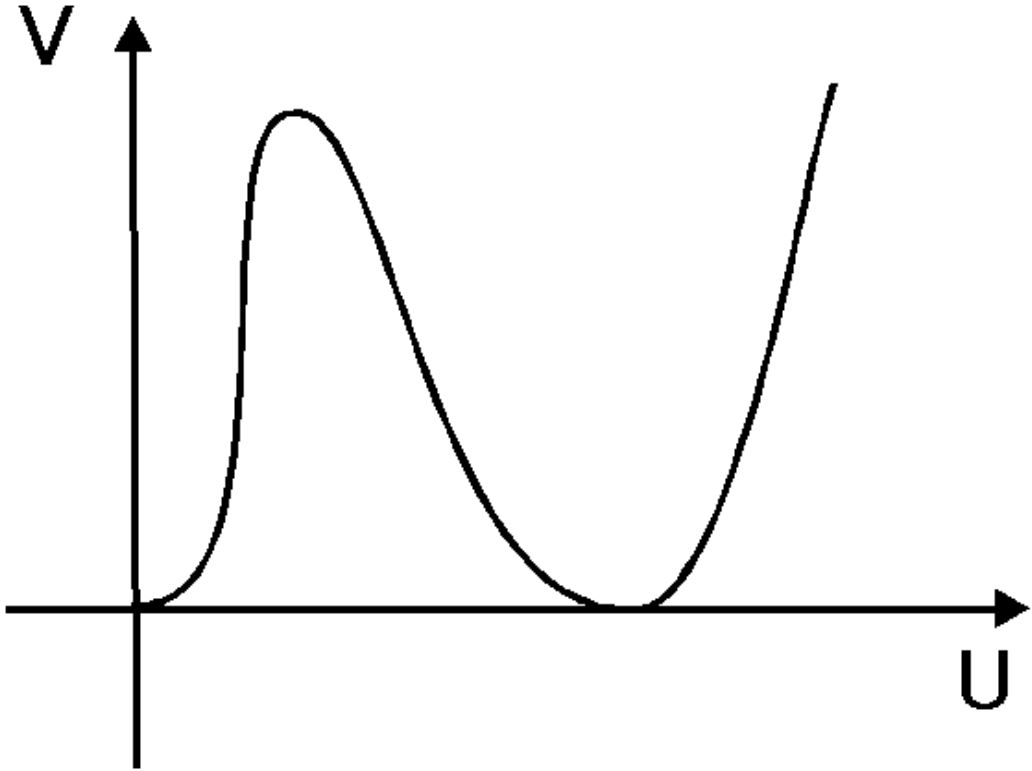}}
  
  \caption{The scalar potential for gaugino condensate(U) for
general supersymmetric Yang-Mills theory.}

  \label{fig:sym1}

\end{figure}

This leads to the effective potential of the form:
\begin{equation}
  \label{sympot}
  V=(K_{U\overline{U}})^{-1}|U|^{2}\left|
   (a+N_{c})+N_{c}log(\frac{U}{\mu^{3}})\right|^{2}
\end{equation}
This scalar potential has the minimum at $U=\mu^{3}e^{-(a+N_{c})/N_{c}}
e^{2\pi i/N_{c}}$ 
which agrees with the analysis of the instanton calculations.
This also agrees with the Witten index analysis, which suggests
that this theory has supersymmetric vacuum with non-zero gaugino
condensation.
One may find an extra minimum at the origin($U=0$), but
because it corresponds to a vanishing point of $K_{U\overline{U}}$
and it also contradicts to other approaches(Witten Index and the instanton
calculation) we may think that
the effective potential is not well defined near the origin
and discard this possibility of finding a trivial minimum at $U=0$.
(See Fig.(\ref{fig:sym1}).)

\newpage

\begin{figure}[hp]
  \epsfxsize=13cm

  \centerline{\epsfbox{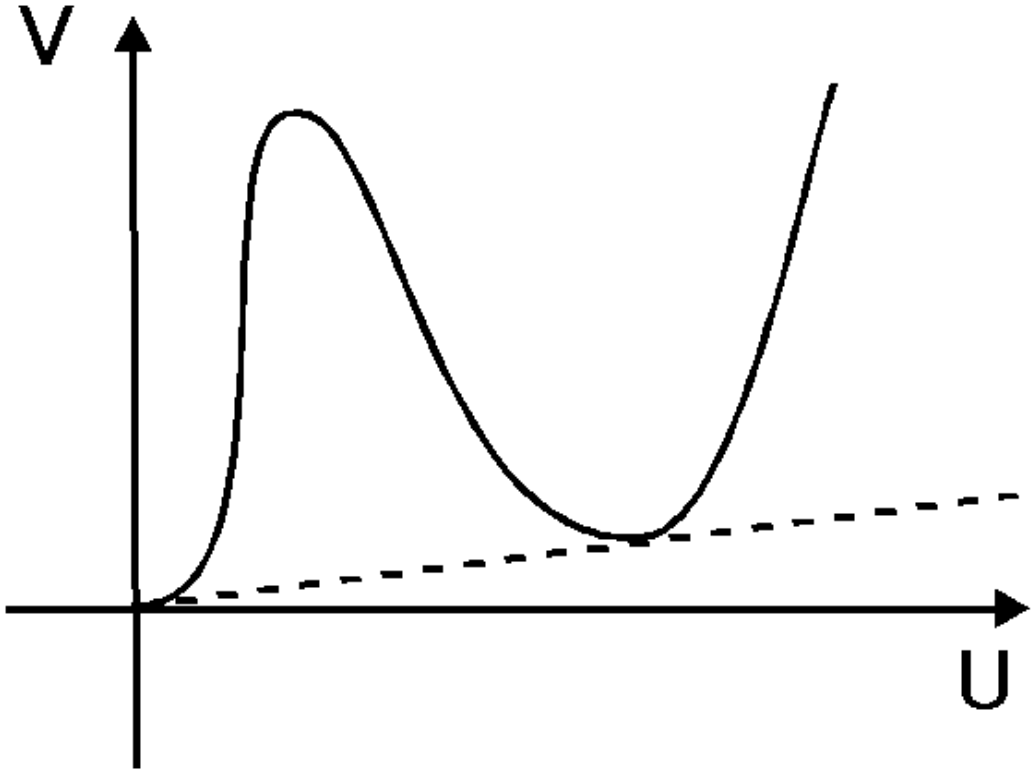}}
  
  \caption{Eq.(3.13)is plotted. It is easy to find that
gaugino condensation breaks supersymmetry once the gauge
coupling constant becomes field dependent.}

 \label{fig:sym2}

\end{figure}

\newpage
Here we also consider an important extension of the model, that is,
 inclusion of field dependent coupling constant.
This extension is well motivated by the superstring effective Lagrangian
or other effective Lagrangian analysis.
Assuming that the inverse of the gauge coupling constant
is given by a function $f(S)$, where $S$ denotes a gauge singlet 
superfield, we obtain an effective Lagrangian of the form:
\begin{equation}
  L=\int d^{4}\theta [K(S,\overline{S})+K(U,\overline{U})]
  +  \int d^{2}\theta [f(S)U+bUlog(\frac{U}{
    \mu^{3}})]+h.c.
\end{equation}
The implication of the field dependent coupling constant is mainly 
discussed by the additional term in the scalar potential, which 
generally implies supersymmetry breaking (or runaway behavior)
that is induced by gaugino condensation.
The explicit form of the scalar potential is:
\begin{equation}
  \label{sym2}
  V=K_{U\overline{U}}^{-1}|f+b+blog(\frac{U}{\mu^{3}})|^{2}+
  K^{-1}_{S\overline{S}}|U|^{2}|\frac{\partial f}{\partial S}|^{2}
\end{equation}
The second term originates from the present gauge singlet superfield
and breaks supersymmetry when gaugino condenses.
If gaugino condensation vanishes at $S\rightarrow\infty$,
this implies a runaway behavior of this potential.
We show the explicit form of this potential in Fig.(\ref{fig:sym2}).

For later convenience, let us examine the most popular example, 
$f(S)=S$ and $K(U,\overline{U})=(U\overline{U})^{1/3}$.
In this case we obtain two solutions.
One is $U=0$ and $S$ arbitrary, and the other is $U\rightarrow 0$
and $S\rightarrow \infty$.

\newpage

\newpage
\subsubsection{Supersymmetric QCD}
\hspace*{\parindent}
According to Witten index theorem, supersymmetry should not
be broken in many theories.
In particular, he has computed the index in pure SU(N) 
Yang-Mills theory and shown that this theory always has at least
N supersymmetric ground state.
This suggests that supersymmetry is  not  broken also in 
supersymmetric QCD theories with any number of massive fields.

However, the index theorem says nothing about the theory with
massless quarks.
This theory has (classical) flat directions along which fields can develop
any value.

In this subsection we review the analysis on the dynamical effects in massless
and massive supersymmetric QCD and show that in supersymmetric
QCD with the number of flavors $N_{f}$, less than $N_{c}$,
a dynamical superpotential is in fact generated.
This potential is given for the composite(confined) fields
for $N_{f}\le N_{c}-1$.
(In this case dynamical  superpotential 
is generated by a  strong coupling effect and 
there remains uncertainty in this calculation at  small field strength.
This uncertainty can be evaded by a peculiar assumption, that
the low-energy theory, which can be thought as a pure Yang-Mills
theory coupled to an R-Axion, has the same characteristics as
the pure Yang-Mills theory.
If this assumption is correct, we can expect that gaugino 
condensation is always non-zero even in the weak coupling limit.)
For $N_{f}=N_{c}-1$, the instanton number constraint does not forbid
the generation of the dynamical superpotential and 
we can  show its generation by an explicit computation.
However for $N_{f}<N_{c}-1$, because of the constraint from the 
instanton number, we should think that instanton is not responsible for
the generation of dynamical superpotential.
(The F-term carries charge 2 under $U(1)_{R'}$ which cancels the
charge -2 from the grassmannian integral. 
On the other hand, from
\begin{equation}
  \partial_{\mu}j^{\mu}_{R}=\frac{1}{32\pi^{2}}2(N_{f}-N_{c})F^{a}
  \tilde{F}^{a}
\end{equation}
we learn that  instanton carries charge $2|N_{c}-N_{f}|$ under 
$U(1)_{R}$ symmetry.
Thus we can see that the instanton-induced
dynamical potential is  allowed only for 
$N_{f}=N_{c}-1$.) 
Because we usually think that 
the instanton effects will dominate the dynamical effects
in the weak coupling region, we think
that the uncertainty does not exist for $N_{f}=N_{c}-1$.
On the other hand, when $N_{f}<N_{c}-1$, we should consider
another dynamical effect.

To obtain dynamical understanding of these theories, many authors
attempted to construct effective Lagrangians to describe the
low-energy dynamics of these theories.
These analysis made it clear that in these theories with
massive quarks, the limit $m_{q}\rightarrow 0$ was likely
to be peculiar.
In particular, it was shown that if a dynamical
superpotential is generated in supersymmetric QCD with
massless quarks by non-perturbative effects, its form is uniquely
determined by the symmetric constraints.
Moreover, if a small mass term is added to this theory,
we can obtain N vacuum states which agrees with the index arguments.
(Of course tree level superpotential can alter the symmetries
of the theory, which have constrained the dynamical superpotential
in massless theories. 
Above we assumed that a small mass term will not change the
qualitative character of the dynamical potentials.)

Let us see more concrete examples. 
Here we mainly follow the paper ref.\cite{ADS} which we regard as
a ``general'' analysis.

By supersymmetric QCD we will mean a supersymmetric theory
with gauge group $SU(N_{c})$ and $N_{f}$ flavors of quarks.
The $N_{f}$ quark flavors correspond to $N_{f}$ chiral fields in the 
$N_{c}$ representation and $N_{f}$ chiral fields in the $\overline{N}_{c}$
representation.
\begin{eqnarray}
  Q^{ir}(i=1,...,N_{c};r=1,...,N_{f}),&&
  \overline{Q}_{ir}(i=1,...,N_{c};r=1,...,N_{f})
\end{eqnarray}
These superfields can be written with component fields as:
\begin{equation}
  \label{compsigma}
  \left\{
    \begin{array}{l}
      Q^{ir}=\phi^{ir}+\theta^{\alpha}\psi_{\alpha}^{ir}+\theta^{2}F^{ir}\\
      \overline{Q}_{ir}=\overline{\phi}_{ir}+\theta_{\alpha}
      \overline{\psi}^{\alpha}_{ir}+\theta^{2}\overline{F}_{ir}
    \end{array}
  \right.
\end{equation}
The gauge fields $A^{a}_{\mu}(a=1,...,N_{c}^{2}-1)$ are included in vector 
multiplets $V^{a}$ accompanied by their super-partners, gauginos $\lambda^{a}$
and auxiliary fields $D^{a}$.
The total theory is given by
\begin{eqnarray}
  \label{lag00}
  L&=&\frac{1}{4g^{2}}\int d^{2}\theta W^{\alpha a}W_{\alpha}^{a}+h.c.
  +\int d^{4}\theta \left[Q^{+}e^{V}Q+\overline{Q}e^{-V}
  \overline{Q}^{+}\right]
\end{eqnarray}
Classically, this theory has a global $U(N_{f})_{Left}\times 
U(N_{f})_{Right}\times U(1)_{R}$ symmetry.
The $U(N_{f})_{Left}\times U(N_{f})_{Right}$ symmetry is just
like that of the ordinary QCD, corresponding to separate rotation of the 
$Q$ and $\overline{Q}$ fields.
The symmetry $U(1)_{R}$ is an R-invariance, a symmetry under which 
the components of a given superfield transform differently.
This corresponds to a rotation of the phases of the grassmannian 
variables $\theta^{\alpha}$,
\begin{equation}
  \label{r}
  \left\{
  \begin{array}{lll}
    \lambda&\rightarrow&e^{i\alpha}\lambda\\
    \psi&\rightarrow&e^{i\alpha}\psi\\
    \overline{\psi}&\rightarrow&e^{i\alpha}\overline{\psi}
  \end{array}
  \right.
\end{equation}
with scalar and vector fields unrotated.
This can be written as:
\begin{equation}
  \left\{
    \begin{array}{lll}
      W_{\alpha}(\theta)&\rightarrow&e^{-i\alpha}W_{\alpha}(\theta 
               e^{i\alpha})\\
      Q(\theta)&\rightarrow&Q(\theta  e^{i\alpha})\\
      \overline{Q}(\theta)&\rightarrow&\overline{Q}(\theta  e^{i\alpha}).
    \end{array}
  \right.
\end{equation}
Just as in the ordinary QCD, some of these symmetries are explicitly 
broken by anomalies.
A simple computation shows that the following symmetry,
which is a combination of the ordinary chiral $U(1)_{A}$ and the $U(1)_{R}$
symmetry, is anomaly-free.
\begin{equation}
  \left\{
    \begin{array}{lll}
      W_{\alpha}(\theta)&\rightarrow&e^{-i\alpha}W_{\alpha}(\theta 
               e^{i\alpha})\\
      Q(\theta)&\rightarrow&e^{i\alpha(N_{c}-N_{f})/N_{f}}
               Q(\theta  e^{i\alpha})\\
      \overline{Q}(\theta)&\rightarrow&e^{i\alpha(N_{c}-N_{f})/N_{f}}
               \overline{Q}(\theta  e^{i\alpha})
    \end{array}
  \right.
\end{equation}
From now on, we call this non-anomalous global symmetry  $U(1)_{R'}$.

It is important to stress that we are not looking for an explicit
breaking of supersymmetry.
Since it is believed that the supersymmetry current has no anomalies,
the effective lagrangian should be supersymmetric and should
be given by superfields.
Its vacua, of course, need not to respect supersymmetry and other
symmetries.
The effective Lagrangian must respect all the (non-anomalous)
symmetries whether the various symmetries
of the theory is broken or not.
Thus the dynamical superpotential which may be generated must be gauge and
$G=SU(N_{f})_{Left}\times SU(N_{f})_{Right}\times U(1)_{V}\times 
U(1)_{R'}$ invariant.
To be gauge and $SU(N_{f})_{Left}\times SU(N_{f})_{Right}\times U(1)_{V}$
invariant, such a F-term must be of the form:
\begin{equation}
  F\sim \int d^{2}\theta f(det_{rr'}\overline{Q}_{ir}Q^{ir'})
\end{equation}
Further requirement from R'-invariance determines the precise form as:
\begin{equation}
   F\sim \int d^{2}\theta (det_{rr'}\overline{Q}_{ir}Q^{ir'})^{
     -1/(N_{c}-N_{f})}
\end{equation}
This term is only meaningful for $N_{f}<N_{c}$.
For $N_{c}=N_{f}$, it is meaningless.
For $N_{f}>N_{c}$, it vanishes identically by a simple symmetry argument.
The coefficient of this F-term should be dimensionful, and must be
given by a power of the dynamically generated scale of the 
theory($\Lambda$), 
which can be related to the scale of gaugino condensation.
\begin{equation}
  \label{np}
  F=const.\Lambda^{\frac{3N_{c}-N_{f}}{N_{c}-N_{f}}}
    \int d^{2}\theta (det_{rr'}\overline{Q}_{ir}Q^{ir'})^{\frac{
        -1}{N_{c}-N_{f}}}
\end{equation}
The question of whether the dynamics indeed generates this term or not
is of course another problem and will be discussed in the following.
(See also ref.\cite{ADS}.)

In $N_{f}=N_{c}-1$, we can show that this term can be constructed 
by an explicit instanton calculation, so the dynamical origin is clear
in this case.

\begin{figure}[ht]
  \epsfxsize=13cm

  \centerline{\epsfbox{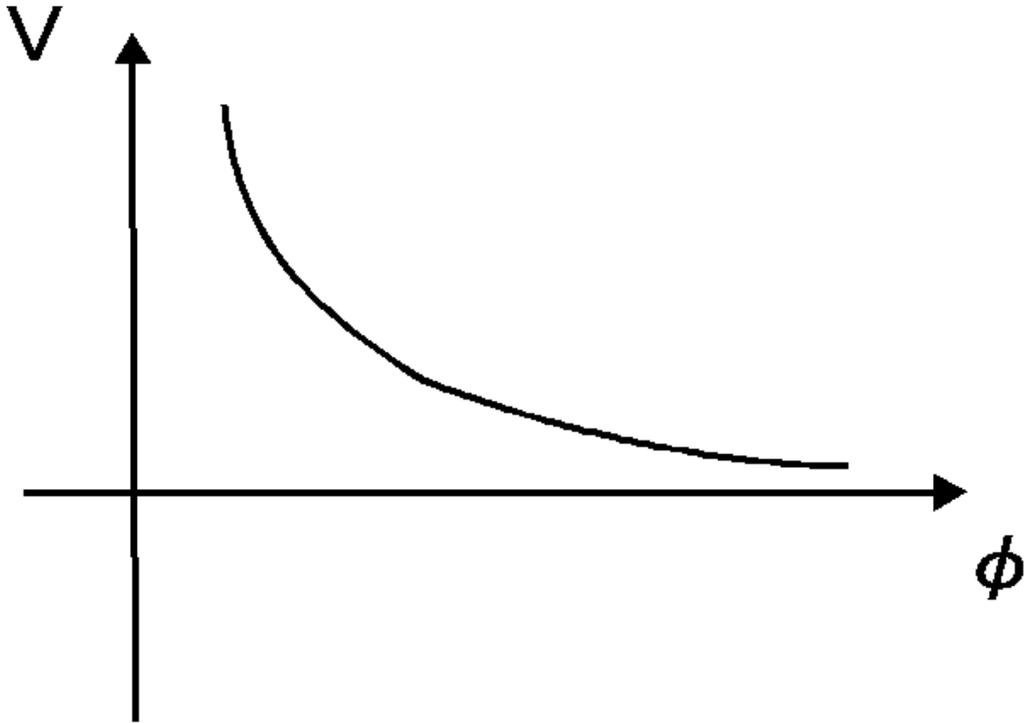}}

 \caption{The flat direction is lifted but there remains
a problematic behavior. This potential presents the so-called
``Runaway potential''.}

 \label{fig:run}

\end{figure}

Since this model has flat directions, it is reasonable to
expect that $Q$ and $\overline{Q}$ may develop their vacuum
expectation values along these directions.
When $N_{f}<N-1$, because instanton  does not allow the 
generation of (\ref{np}), we should explain its generation
from another point of view.
In this case, the gauge group is not completely broken and 
we can expect that the intermediate scale Lagrangian,
which appears after gauge symmetry breaking, is
pure Yang-Mills theory coupled to an axion superfield.
At energies below the symmetry breaking scale, assuming that
the supersymmetry is not explicitly broken by perturbative effects,
we can obtain an effective Lagrangian:

\begin{eqnarray}
  \label{geff1}
  L&=&\int d^{2}\theta \frac{1}{4g^{2}}\left[1+\beta ln
  \left(\frac{\phi}{\Lambda}\right)\right]W^{\alpha}W_{\alpha}
    +h.c.\nonumber\\
    \beta&=&\frac{g^{2}}{32\pi^{2}}N_{f}
\end{eqnarray}
where $\beta$ is determined by R-anomaly.
Eq.(\ref{geff1}) contains a dimension five operator:
\begin{equation}
  \label{2aux}
  L_{AUX}=\frac{\beta g^{2}\lambda\lambda}{v}F_{\phi}+h.c.+
    F^{*}_{\phi}F_{\phi}    
\end{equation}
where $v$ is the symmetry breaking scale of $SU(N_{c})\rightarrow
SU(N_{c}-N_{f})$, which can be identified
with the cut-off scale of this effective Lagrangian.
The field $\phi$ is  considered as an anomaly compensating field for
the non-anomalous R' symmetry of the original Lagrangian.
When $N_{c}=3$ and $N_{f}=1$, this effective Lagrangian is 
explicitly calculated  by integrating the heavy fields\cite{pol}.
This extra term in $L_{AUX}$ is very important when we think
about the dynamical breaking of supersymmetry.
Using the equation of motion for $F_{\phi}$, we obtain the relation
\begin{equation}
  <F^{*}_{\phi}>\sim <\lambda\lambda>/v
\end{equation}
this means that once gaugino condensates in the effective
Yang-Mills theory, supersymmetry should be broken dynamically.
At the same time, the flat direction along $\phi$ is
lifted, but the shape of this potential is rather problematic.
Its minimum lies at infinity, so this potential  is called 
``Run-away'' potential.(See also Fig.(\ref{fig:run}).)
Of course we can use eq.(\ref{sym2}) to obtain dynamical superpotential.

We should not forget that when $\phi$ rolls down to infinitely large
value and run into the weakly coupled region, there remains
uncertainty for $N_{f}<N_{c}-1$.

\newpage

\newpage
So far we have dealt with supersymmetric QCD with massless matters
and we have discussed the generation of non-trivial superpotential
for $N_{f}<N_{c}-1$.
Here we analyze the same theory with (small) mass term for
matter fields.
The supersymmetric mass term is written as:
\begin{equation}
  \int d^{2}\theta \sum^{N_{f}}_{k=1}m_{k}
    \overline{Q}_{ik}Q^{ik}
\end{equation}
For simplicity, we assume that all the masses are equal.
This mass term raises all the flat directions.
According to Witten\cite{WittenIndex}, this theory has al least $N_{c}$
supersymmetric vacua and supersymmetry is not  broken by
any dynamical effects.
To show how this vacua are realized in the dynamical phase,
we would like to discuss the symmetries of the massive QCD.

If this theory does not contain any mass term, the symmetry is(see the 
previous section):
\begin{equation}
  G=SU(N_{f})_{Left}\times SU(N_{f})_{Right}\times U(1)_{V}\times U(1)_{R}
\end{equation}
This symmetry is broken by the mass term.
The continuous part of the remaining symmetry is:
\begin{equation}
  SU(N_{f})\times U(1)_{V}
\end{equation}
where $SU(N_{f})$ is the vector subgroup of $SU(N_{f})_{Left}\times 
SU(N_{f})_{Right}$.
Besides this continuous symmetry, this theory has discrete $Z_{2N_{c}}$
symmetry:
\begin{eqnarray}
  W(\theta)&\rightarrow& e^{-i\pi n/N_{c}}W(\theta e^{i\pi n/N_{c}})\nonumber\\
  Q(\theta)&\rightarrow& e^{-i\pi n/N_{c}}Q(\theta e^{i\pi n/N_{c}})\nonumber\\
  \overline{Q}(\theta)&\rightarrow& 
  e^{-i\pi n/N_{c}}\overline{Q}(\theta e^{i\pi n/N_{c}})\nonumber\\
  n&=&1,...,2N_{c}
\end{eqnarray}

\begin{figure}[hb]
    \epsfxsize=13cm

  \centerline{\epsfbox{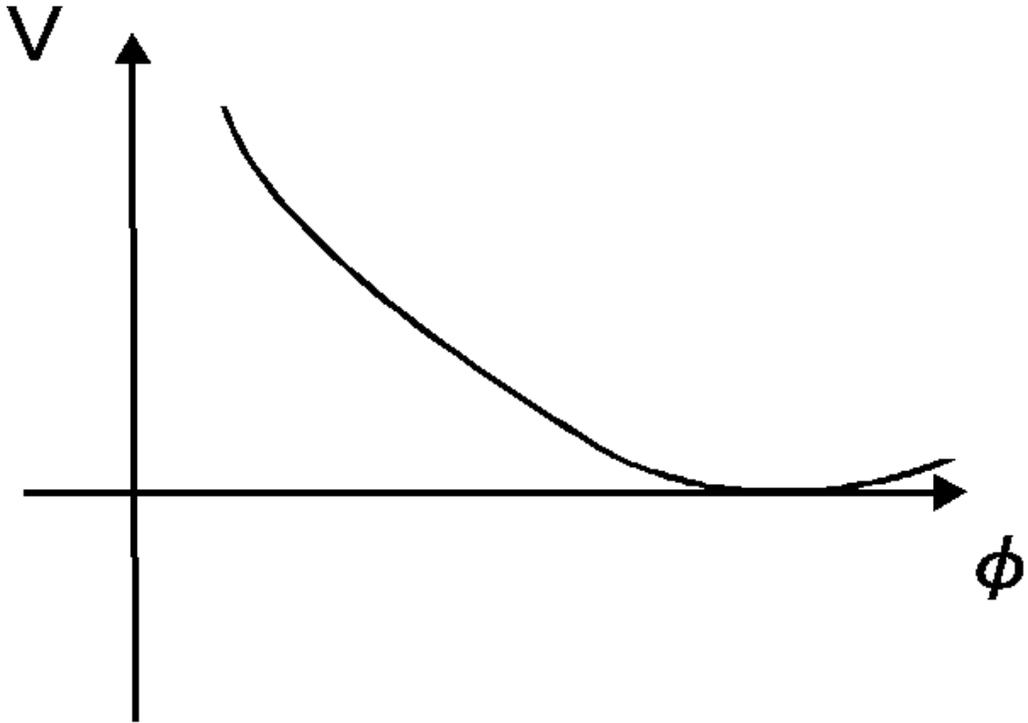}}

 \caption{The scalar potential for SQCD with small mass term.
The asymptotic behavior is changed.
No runaway is observed and supersymmetry is kept unbroken.}

 \label{fig:massive}

\end{figure}

Witten has suggested that the $N_{c}$ vacua counted by index argument
might be associated with the spontaneous breaking of this symmetry
to a $Z_{2}$ subgroup.
As far as the quark mass is small and can be considered as a small
perturbation, one can show that the dynamical superpotential
is still generated.
The full superpotential is given by:
\begin{equation}
  W_{dyn}=const.\Lambda^{\frac{3N_{c}-N_{f}}{N_{c}-N_{f}}}
    (det\overline{Q}Q))^{\frac{-1}{N_{c}-N_{f}}}
    +m\overline{Q}Q
\end{equation}
This superpotential has N supersymmetric vacua of the form:
\begin{eqnarray}
  <\overline{Q}Q>&=&\Lambda^{3-\frac{N_{f}}{N_{c}}}
  \left(\frac{const.}{m(N_{c}-N_{f})}\right)^{-1+\frac{N_{f}}{N_{c}}}
    e^{2\pi in/N_{c}}
    \nonumber\\
    n&=&1,...,N_{c}
\end{eqnarray}
This means that supersymmetric equation has exactly $N_{c}$
different solution.
The runaway potential is thus stabilized and supersymmetric
vacuum is now well defined.
(See Fig.(\ref{fig:massive})

\newpage

\newpage
\subsection{Large N expansion of global supersymmetric models}
\hspace*{\parindent}
In this section, we will discuss dynamical properties
of global supersymmetric gauge theories in terms of large N 
expansion.
This method can easily be extended to supergravity models
\cite{matsuda_QCD,matsuda_sugra}.

\subsubsection{Gaugino condensation in supersymmetric QCD}
\hspace*{\parindent}
Gaugino condensation in supersymmetric gauge theories has been extensively
studied by many authors both in global\cite{rep} and 
local\cite{leff} theories.
In this section we examine the vacuum structures of 
Supersymmetric QCD(SQCD) theories with $N_{f}<N_{c}-1$
by using the Nambu-Jona-Lasinio method with large $N$ expansion.
We follow ref.\cite{ADS} in deriving the intermediate 
effective Lagrangian.

The results presented below nicely agree with the previous studies which are 
given by the instanton calculation or effective Lagrangian analysis.

Our starting point is a Lagrangian with a gauge group $SU(N_{c})$ with 
$N_{f}$ flavors of quarks.
These superfields can be written with component fields as (The following
 Lagrangian
is the same as what was analyzed in the previous section.):
\begin{equation}
  \label{component}
  \left\{
    \begin{array}{l}
      Q^{ir}=\phi^{ir}+\theta^{\alpha}\psi_{\alpha}^{ir}+\theta^{2}F^{ir}\\
      \overline{Q}_{ir}=\overline{\phi}_{ir}+\theta_{\alpha}
      \overline{\psi}^{\alpha}_{ir}+\theta^{2}\overline{F}_{ir}
    \end{array}
  \right.
\end{equation}
The gauge fields $A^{a}_{\mu}(a=1,...,N_{c}^{2}-1)$ are included in vector 
multiplets $V^{a}$ accompanied by their super-partners, gauginos $\lambda^{a}$
and auxiliary fields $D^{a}$.
The total theory is given by
\begin{eqnarray}
  \label{lag0}
  L&=&\frac{1}{4g^{2}}\int d^{2}\theta W^{\alpha a}W_{\alpha}^{a}+h.c.
  +\int d^{4}\theta \left[Q^{+}e^{V}Q+\overline{Q}e^{-V}
  \overline{Q}^{+}\right].
\end{eqnarray}
The symmetries of this model are already discussed in the previous
section.

Since this model has flat directions, it is reasonable to
expect that $Q$ and $\overline{Q}$ may develop their vacuum
expectation values along these directions.
If $N_{f}<N-1$, the gauge group is not completely broken.
Moreover, we can see that instantons cannot generate a superpotential
in this case, so we think that considering another type of 
non-perturbative effects 
in this model seems very important.

For simplicity, here we consider the case: $SU(N_{c})$ gauge group is
broken to $SU(N_{c}-N_{f})$.
The low-energy theory, where gauge interaction is still unconfined,
 consists of two parts: Kinetic terms
for the unbroken pure $SU(N_{c}-N_{f})$ 
gauge interaction and one for the 
massless chiral field.
In addition to these terms, we should include higher dimensional
operator.
A dimension-five operator, in general, is generated at 
one-loop level\cite{pol}.
As we have stated in the previous section, this can be obtained also 
from the renormalization of the effective
coupling\cite{rep}:
\begin{equation}
  \label{geff}
  L=\int d^{2}\theta  \frac{1}{4g^{2}}\left[1+\frac{g^{2}}{32\pi^{2}}N_{f}ln
  \left(\frac{\phi}{\Lambda}\right)\right]W^{\alpha}W_{\alpha}
  +h.c.
\end{equation}
where we can think that the renormalization of the effective
gauge coupling constant is now field dependent:
\begin{equation}
  g^{-2}(\phi)\equiv 
g^{-2}[1+\frac{g^{2}}{32\pi^{2}}N_{f}ln(\frac{\phi}{\Lambda})]
\end{equation}
Of course, this term itself is not five dimensional.
Redefining the field as $\phi=<v>+\phi'$, this term
produces a dimension five operator, namely $\sim\frac{\phi'}{v}W^{2}$.
Here $v$ can be regarded as a symmetry breaking scale and can be
treated as the cut-off scale of the low energy
effective theory($v\sim \Lambda$).
In general, each $Q$ can develop different values, but here,
for simplicity,
we assume every $Q$ gains the same classical value.
$\phi$ must be chosen to be invariant under all 
symmetries except for $U(1)_{R'}$.
Detailed arguments on such a field dependence of
coupling constant are given in ref.\cite{rep} and the references therein.
The non-anomalous R'-symmetry of the original theory must be realized
in the effective low-energy Lagrangian by the shift
induced by $\phi$.
That determines the R'-charge of $\phi$ to be $(N_{c}-N_{f})/N_{f}$.

For simplicity, we consider a generalized form
\begin{equation}
  \label{1qcd}
  L=\int d^{2}\theta \frac{1}{4}f(\phi)W^{\alpha}W_{\alpha}+h.c.
  +\int d^{4}\theta \phi^{*}\phi
\end{equation}
where $f(\phi)$ is the field dependent coupling constant.
\begin{eqnarray}
  \label{1a}
  f(\phi)&=&\frac{1}{g_{0}^{2}}+\beta log\left(\frac{\phi}{\Lambda}
  \right)\nonumber\\
\end{eqnarray}
Here $\beta$ is a constant chosen to realize the anomaly free
(mixed) R'-symmetry of the original Lagrangian.
In our case, we take $\beta=\frac{N_{f}}{32\pi^{2}}$.

Generally, the kinetic term for the axionic superfield,
$\phi\phi^{*}$ in eq.(\ref{1qcd}), is not calculable and one may
expect it should have some other  complicated form.
However, as far as $K(\phi,\phi^{*})$ is written by a 
function of the form $f(\phi\phi^{*})$, the essential
features are not changed.
In that case, the auxiliary part of the kinetic term is
changed to: $F_{\phi}F_{\phi}^{*}\rightarrow f''F_{\phi}F_{\phi}^{*}$.
Here, for convenience, we consider the simplest choice.
The gauge group of the low energy theory is $SU(N_{c}-N_{f})$.
What we are concerned with is the auxiliary part of this Lagrangian:
\begin{equation}
  \label{2}
  L_{AUX}=\frac{\beta g^{2}\lambda\lambda}{v}F_{\phi}+h.c.+
    F^{*}_{\phi}F_{\phi}    
\end{equation}
(This term can be derived by a direct integration of massive
fields\cite{pol}.)
We can simply assume that the cut-off scale of this effective
Lagrangian is $v$.
The factor of $g^{2}$ appears because we have rescaled gaugino fields
to have canonical kinetic terms.
The equation of motion for $F_{\phi}$ is:
\begin{eqnarray}
  \label{2a}
  \frac{\partial L}{\partial F_{\phi}}&=&\frac{\beta g^{2}}{v}\lambda\lambda
  +F_{\phi}^{*}\nonumber\\
  &=&0
\end{eqnarray}
This equation means that $<\lambda\lambda>$ is proportional to
$F_{\phi}$. 
We can think that $<\lambda\lambda>$ is
the order parameter for the supersymmetry breaking.
Using the tadpole method\cite{tad} we can derive a gap equation directly
from (\ref{2a}).
\begin{eqnarray}
  \label{2b}
  F^{*}_{\phi}&\times&\left(1-4G^{2}\int \frac{d^{4}p}{(2\pi)^{4}}
   \frac{1}{p^{2}
    +m_{\lambda}^{2}}\right)=0\nonumber\\
  &&\left\{
    \begin{array}{l}
      G^{2}=\frac{\beta^{2} g^{4}}{v^{2}}n_{g}\\
      m_{\lambda}^{2}=\frac{|F_{\phi}|^{2}g^{4}\beta^{2}}{v^{2}}        
\end{array}
  \right.
\end{eqnarray}
where $\beta$ is proportional to $N_{f}$ and $n_{g}$ is the
dimension of the low energy gauge group.
(In this model $n_{g}$ is defined as $n_{g}=(N_{c}-N_{f})^{2}-1$)

\begin{figure}[hp]
  \epsfxsize=13cm

  \centerline{\epsfbox{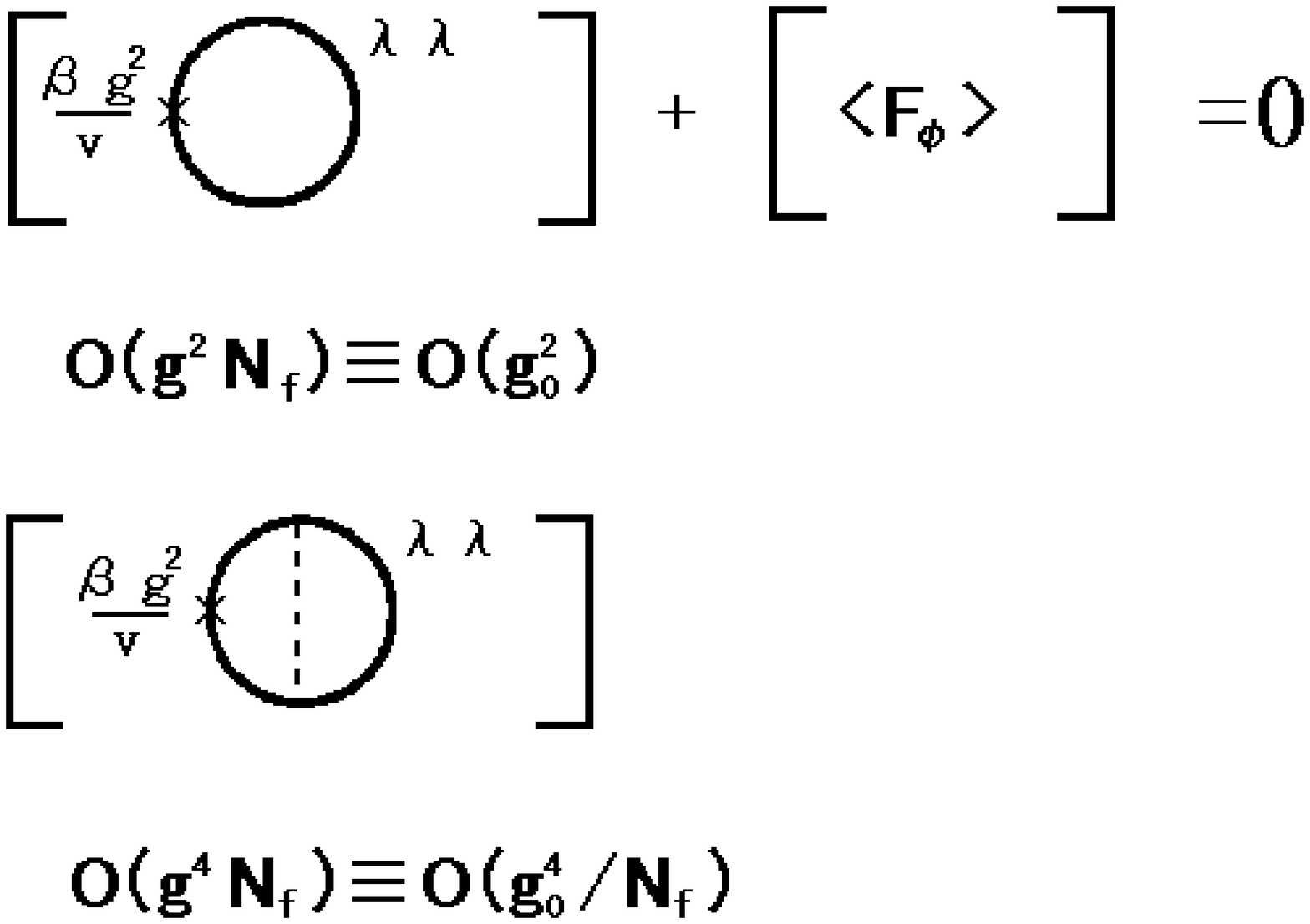}}

 \caption{The  gap equation (3.40) is written 
in a diagrammatic form.
One of the typical  correction is shown in the second diagram.}

 \label{fig:largeN_QCD}

\end{figure}

Taking the limit $N_{f}\rightarrow\infty$, the above equation
becomes a good approximation in a sense of large N expansion. 
(See fig.(\ref{fig:largeN_QCD}))

Of course one should be able to derive (\ref{2b}) by explicit calculation
of 1-loop effective potential.
But it will be very difficult because in calculating the 
explicit 1-loop effective potential we should include the 
superpartner of gaugino condensation(may be a glueball),
which makes this analysis much more complicated.

\begin{figure}[hp]
  \epsfxsize=13cm

  \centerline{\epsfbox{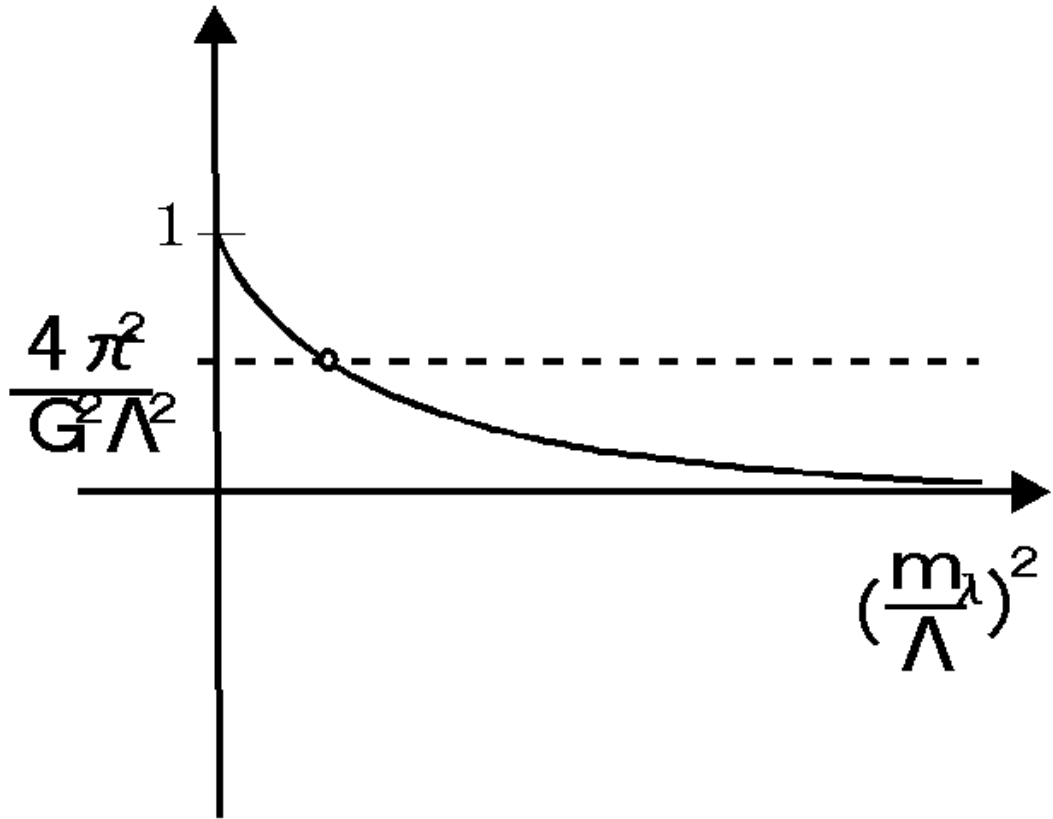}}

 \caption{Left hand side of eq.(3.42)(dotted line)
and right hand side(solid line) is plotted.
If the four-fermi coupling is strong enough, non-trivial
solution appears.}

 \label{fig:gapsol}

\end{figure}

Let us examine the solution of this gap-equation.
After integration we can rewrite it in a simple form.
\begin{equation}
  \label{gapsol}
  \frac{4\pi^{2}}{G^{2}\Lambda^{2}}=1-\left(\frac{m_{\lambda}^{2}}
  {\Lambda^{2}}\right)ln\left(1+\frac{\Lambda^{2}}{m_{\lambda}^{2}}
  \right)
\end{equation}
(See also Fig.(\ref{fig:gapsol}).)

\begin{figure}[hp]
  \epsfxsize=13cm

  \centerline{\epsfbox{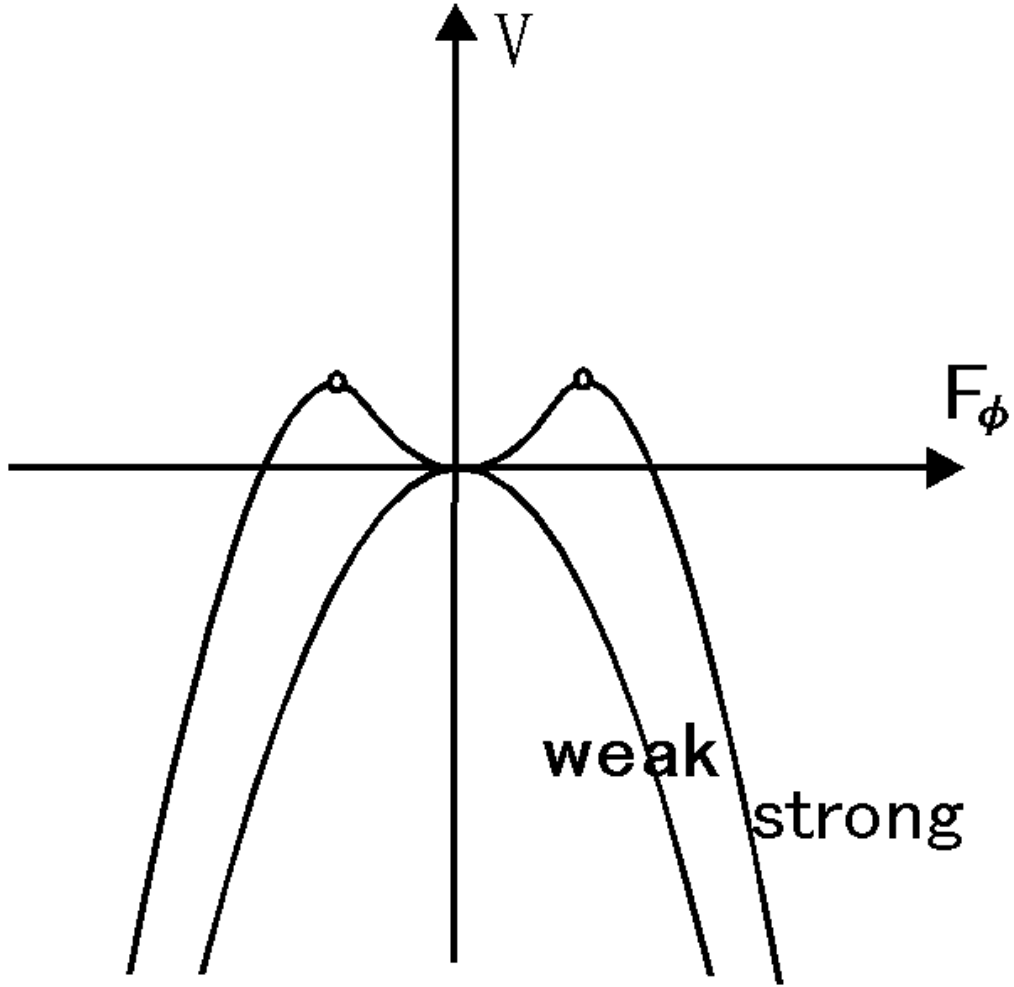}}

 \caption{The relation between $F_{\phi}$ and the scalar potential is plotted
for the weak and the strong coupling phases of
massless SQCD with $N_{f}<N_{c}-1$.
The explicit form of this potential is not calculated explicitly, but
the stationary point of this potential is already calculated in (3.41).
The solution for the gap-equation always lies at the stationary point
of this potential.
Because $F_{\phi}$ is an auxiliary field, the form of the potential
is not important.}

 \label{fig:1loopmassless}

\end{figure}

\begin{figure}[hp]
  \epsfxsize=13cm

  \centerline{\epsfbox{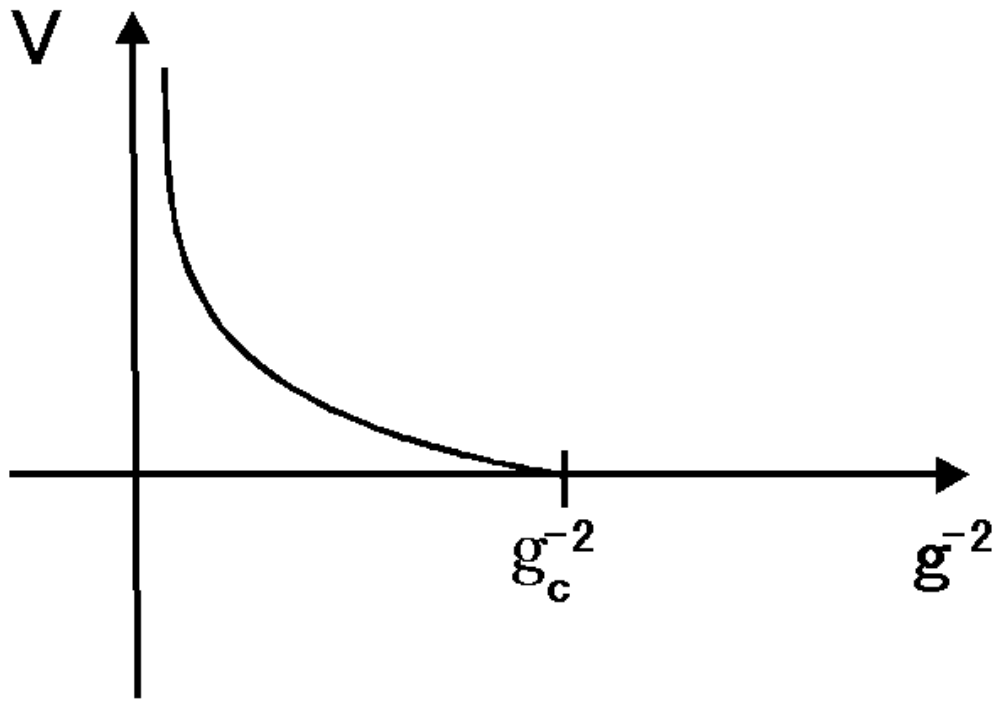}}

 \caption{The explicit 1-loop potential for massless SQCD
with $N_{f}<N_{c}-1$.
Asymptotic behavior is different from the effective Lagrangian
analysis.
We are not sure whether this difference comes from
an artificial reason or there is really a phase transition
for gaugino condensation.
In pure supersymmetric Yang-Mills theory, we have good reasons
to believe that there is no phase transition  and gaugino condensates
for any value of $g$, but in this case it is very difficult
to examine.}

 \label{fig:runalt}

\end{figure}

\begin{figure}[hp]
  \epsfxsize=13cm

  \centerline{\epsfbox{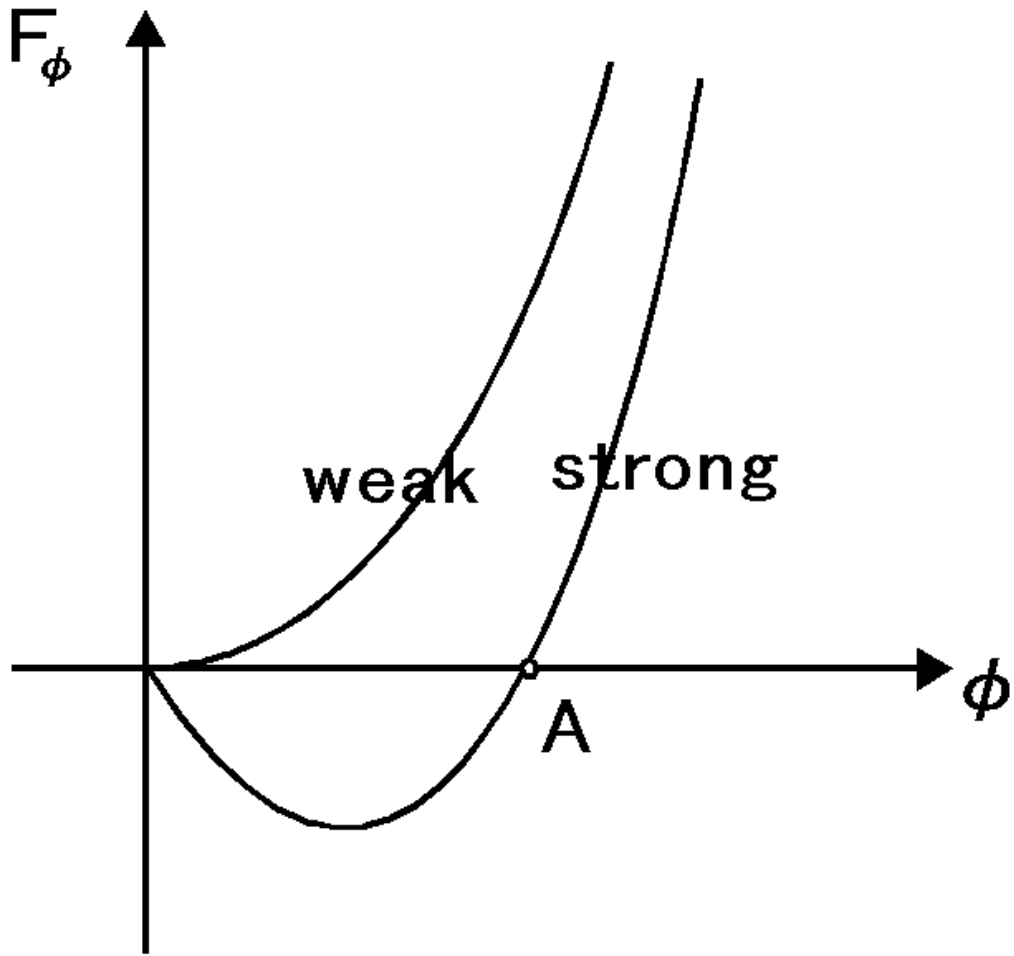}}

 \caption{The relation between $F_{\phi}$ and $\phi$ for SQCD with small 
mass term
is plotted.
It is obvious that we can find a supersymmetric vacuum state
with vanishing $F_{\phi}$.}

 \label{fig:massivef}

\end{figure}

\begin{figure}[hp]
  \epsfxsize=13cm

  \centerline{\epsfbox{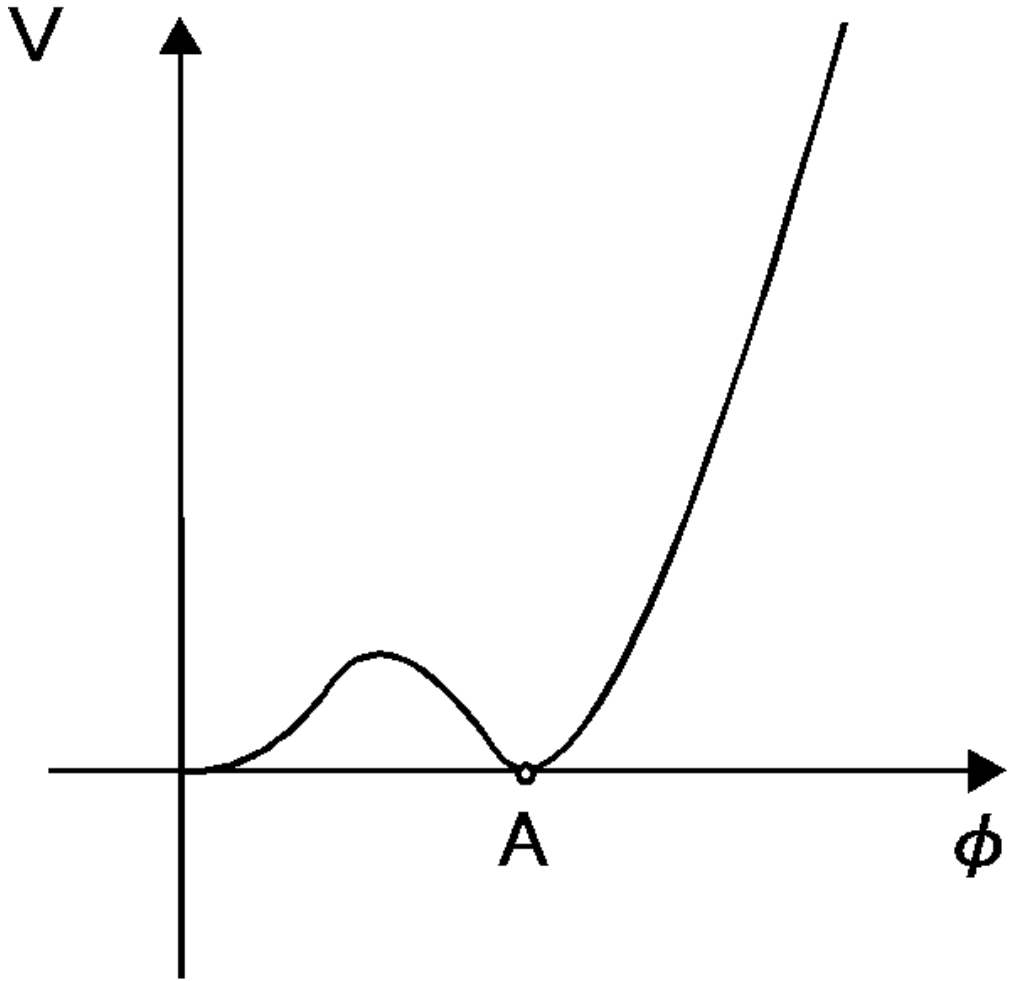}}

 \caption{The explicit form of scalar potential is plotted
for the weakly and the strongly coupled phases.
We can find a ``stable'' supersymmetric vacuum state.}

 \label{fig:massivev}

\end{figure}

\begin{figure}[hp]
  \epsfxsize=13cm

  \centerline{\epsfbox{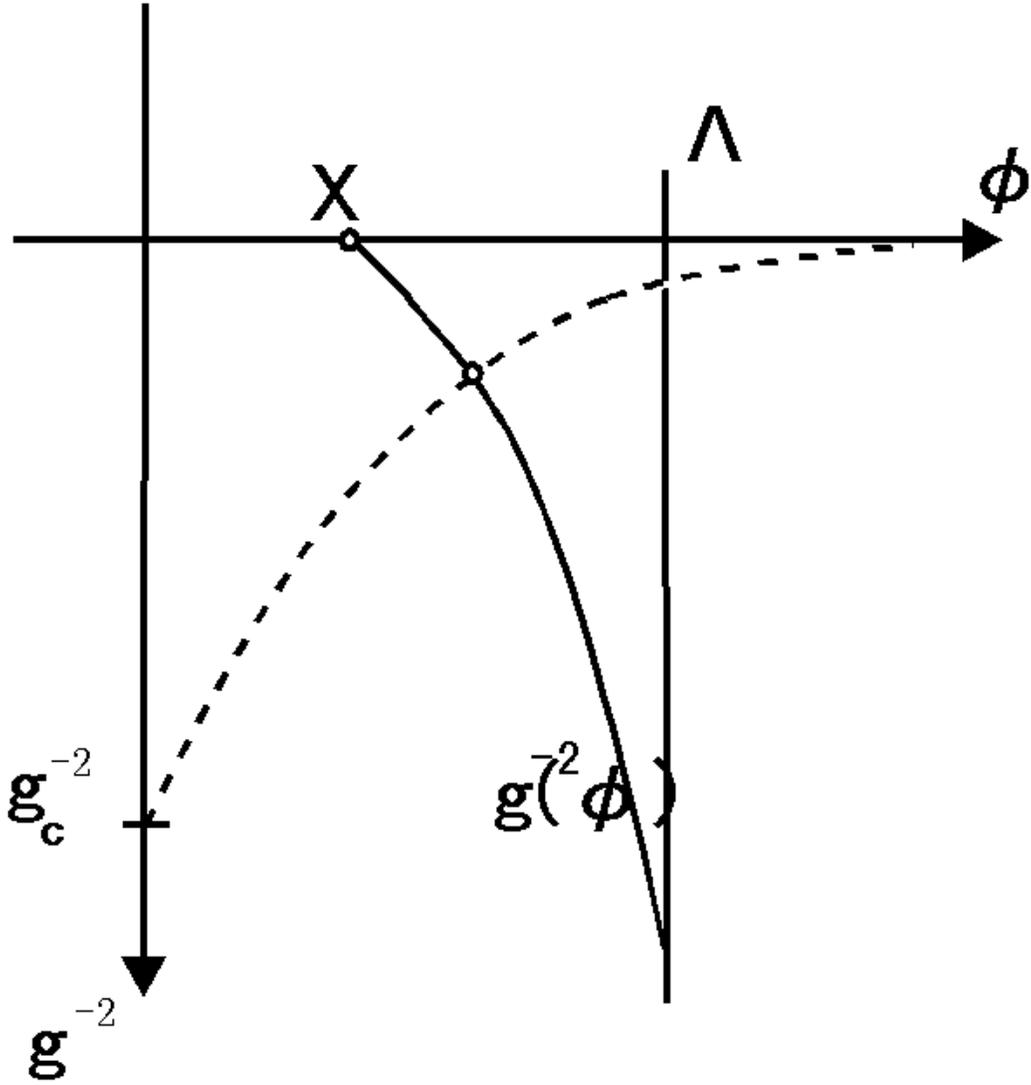}}

 \caption{The dotted line presents the  solution A in 
Fig.9 and Fig.10 for arbitrary $g$.
This line is obtained by solving the equation (3.45) 
with the assumption that $g$ does not depend on $\phi$.
(See also eq.(3.42) and fig.6)
 The solid line shows the relation
between $g$ and $\phi$.
To find a true vacuum, we should find a point at which both
of these two lines meet.
Seeing this graph, we can easily understand why  no phase 
transition is observed in massive SQCD.
In general, we  tend to think that the vacuum lies at
$X$, which corresponds to a solution for $f(\varphi)=0$.}

 \label{fig:massivesol}

\end{figure}

In the strong coupling region, this equation can have non-trivial
solution.
The explicit form of the scalar potential is shown in 
Fig.(\ref{fig:1loopmassless}) and Fig.(\ref{fig:runalt}).
(Here we ignore the trivial solution $F_{\phi}=0$. 
We should note that such a  solution does exist also in the effective 
 Lagrangian analysis
of pure Yang-Mills theory(see section 3.1), but it was neglected from several
reasons.)
Let us examine the behavior of this non-trivial
solution.
In pure supersymmetric Yang-Mills theories, gaugino condensation
is observed even in the weak coupling region because of the
instanton calculation
and Witten index argument that suggests
the invariance of Witten index under the deformation of 
coupling constants\cite{WittenIndex}.
If we believe that the characteristics of the low energy
Lagrangian of massless SQCD is 
also similar to pure SYM, the weak coupling region should be
lifted by gaugino condensation effect.
On the other hand, if we believe that non-compactness of the moduli
space is crucial 
and believe that gaugino condensation
should vanish in the weak coupling region, we can think that the potential 
represented in Fig.(\ref{fig:runalt}) 
is reliable and potential is flat in the weak coupling
region.
We cannot make definite answer to this question, but some suggestive
arguments can be given by adding a small mass term for the field
$\phi$.
\begin{equation}
  L^{add}_{mass}=\frac{1}{2}\epsilon \phi^{2}
\end{equation}
Existence of this term suggests that the moduli space is now compact.
The resulting gap equation is drastically changed.
We can naturally set $F$ components to vanish, and the equation turns out
to be a non-trivial equation for ``$\phi$''.
Relevant terms are:
\begin{equation}
  \label{3}
  L_{AUX}=\left(\frac{\beta g^{2}}{v}\lambda\lambda+\epsilon 
  \phi\right)F^{*}_{\phi}
  +h.c.+F^{*}_{\phi}F_{\phi}
\end{equation}
The equation of motion for $F_{\phi}$ suggests that $<\lambda\lambda>$
is now proportional to $\phi$ and no longer an order parameter for
the supersymmetry breaking.
The gap equation is given by:
\begin{eqnarray}
  \label{3a}
  \epsilon \phi&\times&\left(
  1-4G^{2}
  \int \frac{d^{4}p}{(2\pi)^{4}}
  \frac{1}{p^{2}+m_{\lambda}^{2}}\right)=0\nonumber\\
    &&\left\{
    \begin{array}{l}
    G^{2}=\frac{\beta^{2} g^{4}}{v^{2}}n_{g}\\
    m_{\lambda}^{2}=\frac{\epsilon^{2} g^{4}\beta^{2} |\phi|^{2}}{v^{2}}
    \end{array}
    \right.
\end{eqnarray}
In general, this equation has a solution $m_{\lambda}=const.$
(see Fig.(\ref{fig:massivef})and (\ref{fig:massivev})) which 
does {\it not} break supersymmetry, and does not 
change Witten index for any(non-zero) value of $\epsilon$ and $g_{0}$.
In this case, the potential energy is always $0$ for any value of $g$.
Because the moduli space is compact and Witten index is
well defined in this case, 
it is conceivable that there is no phase
transition for  gaugino condensation.
(See Fig.(\ref{fig:massivesol}).)


Here, we also comment on a peculiar properties of this model.
As we have seen in the exactly calculable models,
the dynamical mass is stable against the soft gaugino mass
in the strong coupling region(i.e. when the gap-equation develops
non-trivial solution.).
This is precisely true in the large N limit, but we are not
sure whether this property remains true also in the phenomenological
models.

\newpage

\newpage

\newpage

\newpage

\subsection{Summary of section 3}
\hspace*{\parindent}
In the first half of this section we have reviewed ref.\cite{ADS}
as an  introduction to the dynamical analysis of supersymmetric QCD.
In the latter half, we considered  a new method for the analysis of 
gaugino condensation and the generation of the non-perturbative potential.
Our analyses are almost consistent with the previous ones.
Moreover, a non-trivial assumption, that was made in \cite{ADS},
is examined. 
In the large N limit we have shown 
that the phase transition of 
gaugino condensation does not occur in massive case but does occur in 
massless case.

\newpage
\section{Dynamical analysis in supergravity theories}
\hspace*{\parindent}

In general, one may think there are four different ways of 
introducing gaugino condensates into supergravity.

First, in the component Lagrangian method\cite{First}, one takes the
 standard Lagrangian
of supersymmetric Yang-Mills theory coupled to supergravity and 
 replaces the gaugino bilinear
with a constant of the order of the condensation scale $\mu^{3}$.
Such a procedure has the drawback of discarding the back-reaction of 
other fields, hence one cannot determine in this way whether the 
condensate really forms.
The formation of the condensate and its magnitude here are simply 
assumed implicitly relying on the observation made in the global version
of the model.
However, as we have assumed that gravitational corrections
can play an important role, the internal consistency of this approach is 
not clear.(There is no reason to believe that dynamical properties
of the global supersymmetric models are not changed in supergravity models.)

Second, a refinement of this approach is considered\cite{DRSW}.
Taking into account a possible
dependence of condensate on some fields leads to the superpotential 
method.
Using these one can then construct the gaugino induced non-perturbative
corrections to the original superpotential of the model.
For example, the belief that the condensate dissolves 
in the weak coupling limit leads to a superpotential
which decays exponentially with the increasing value of the
dilaton field.
One may then search for minima of the effective theory and determine
the true value of the gaugino condensate that leads to 
the supersymmetry breaking in supergravity theories.

The third method is the effective Lagrangian approach\cite{casas1}.
Generalizing the global effective potential written by the 
composite superfields, one can obtain the supergravitational
version of the effective potential which is of course consistent
to the global theory in the $M_{p}\rightarrow\infty$ limit.
In supergravity models the effective scalar potential is:
\begin{equation}
  V=g_{ij^{*}}F^{i}F^{j^{*}}-3e^{G}
\end{equation}
where $G=K+log(|W|^{2})$ and 
$g_{ij^{*}}=\partial^{2}K/\partial z^{i}\partial z^{j^{*}}$.

The fourth method is the Nambu-Jona-Lasinio like approach
first developed by G.G.Ross et al\cite{ross}.
One of the major  advantages of this method is the clear 
relation between  the
constituent Lagrangian and the effective Lagrangian with gaugino 
condensation.
The driving force, that makes gaugino condensate as the coupling
becomes strong, is now clear.
The stability of the moduli potential is also modified,
which we think as  another  advantage of this method.

\subsection{Review of the general analysis}
\hspace*{\parindent}

There has recently been considerable attention focused on the study of 
supersymmetric models of elementary particle interactions.
This is especially true in the context of grand unification theories,
where remarkable studies have been done in the hope of solving the
gauge hierarchy problem or unifying the gravitational interaction
within the superstring formalism.
Supersymmetric extension of the gravity(supergravity) seems necessary 
when one  introduces soft breaking terms and  
makes the cosmological constant vanish at the same time.
In supergravity models,  spontaneous breaking of local supersymmetry
or super-Higgs mechanism may generate soft supersymmetry breaking
terms that allow to fulfill such phenomenological requirements.
However, the super-Higgs mechanism implies the existence of a 
supergravity breaking scale, intermediate between the Planck scale($M_{p}$)
and the weak scale($M_{W}$).
The intermediate scale is expected to be of O($10^{13}$Gev).
Here we expect that this intermediate scale is implemented by the 
mechanism of gaugino condensation in the hidden sector which couples
to the visible sector by gravitational interactions.
The effective action for 
gaugino condensation is well studied by many authors\cite{rep,leff}.

Before discussing the large N expansion of supergravity,
we will review the general approaches to the dynamical 
properties.
In this section, we mainly follow ref.\cite{leff}
and study gaugino condensation in the hidden sector in a modular
invariant way.
Gaugino condensation  is, in general, believed to be a potential source of
hierarchical supersymmetry breaking and the source of a non-trivial potential
for the dilaton $S$, whose real part corresponds to the tree level
gauge coupling constant.
In the effective Lagrangian approach, however, we cannot find a 
stable potential for dilaton without introducing multiple
gaugino condensations.
However, if we include  hidden matter fields with multiple
hidden gauge groups, we can obtain a reasonable value for $S$
and soft breaking terms.

First we consider gaugino condensation without matter fields.
the strategy is as follows.
\
\begin{itemize}
\item Derive the general form of the corresponding scalar potential and 
  the minimization conditions.
\item Consider in a separate way the case of one, two or more
  gaugino condensations.
\end{itemize}
(Multiple gaugino condensation can lead to a stable dilaton potential,
but to produce realistic soft terms we need some extensions, for
example, inclusion of hidden matter fields.
In this sense, hidden matter fields are necessary for realistic model 
building.) 

The process of gaugino condensation in the context of a pure 
Yang-Mills N=1 supergravity theory has been pretty well 
understood for a long time.
It can be described conveniently by an effective superpotential $W^{np}(U)$
of the chiral composite superfield $U$ whose scalar component
corresponds to the gaugino composite bilinear field.
Assuming that the form of $W^{np}$ is the same as pure
supersymmetric Yang-Mills theory, $W^{np}$ reads:
\begin{equation}
  \label{np1}
  W^{np}=aU(f+\frac{2}{3}\beta logU)
\end{equation}
Here $a$ is some constant.
In this case $f$ would be modular dependent and may contain the
threshold corrections from underlying string theory.
As has been demonstrated in refs.[\cite{carlos7,carlos11,carlos12}], 
it is equivalent to work
with the explicit form of $W^{np}$(\ref{np1}) or with the
resulting superpotential after substituting the minimum condition
$\partial W^{np}/\partial U =0$:
\begin{eqnarray}
  W^{np}&=&d e^{\frac{3}{2\beta}f}\nonumber\\
  &=&d \frac{e^{\frac{-3kS}{2\beta}}}{\Pi^{3}_{i=1}[\eta(T_{i})]^{
      2-\frac{3k}{4\pi^{2}\beta}\delta^{GS}_{i}}}
\end{eqnarray}
where $d=-\beta/6e$.
If the gauge group is not simple, i.e.$ G=\Pi_{a}G_{a}$, then 
$W^{np}=\sum_{a}W^{np}_{a}$.
Now the scalar potential is
given by:
\begin{eqnarray}
  V&=&\frac{2}{Y\Pi^{3}_{i=1}T_{Ri}}
    \left[ 
      |YW_{S}-W|^{2}\frac{}{}\right.\nonumber\\
      &&+\sum^{3}_{i=1}\frac{Y}{Y+\frac{1}{4\pi^{2}}\delta^{GS}_{i}}
      \left|
        (W+\frac{1}{4\pi^{2}}\delta^{GS}{i}W_{S})
        -\frac{1}{2}T_{Ri}W_{T_{i}}\right|^{2}\nonumber\\
       && \left. -3|W|^{2}\right]
\end{eqnarray}
Using this scalar potential and numerical methods,
the property of the vacuum state was studied.
Here we only show the results of the previous papers\cite{leff}

\begin{itemize}
\item With simple gauge group\\
  In this case, we cannot find the stable vacuum state.
  The asymptotic minimum appears at infinity, that is called
  ``runaway vacuum''.
\item With multiple gauge groups\\
  If $k_{a}$ is different from each other, we can find a stable minimum.
  However, such a vacuum does not have  phenomenologically
  acceptable values of moduli fields.
\item With hidden matters with multiple gauge groups\\
  Including multiple hidden matter fields, we can obtain a
  phenomenological vacuum state.
  But one should not think that the inclusion of hidden matter is 
  essential, because such an extra parameter region generally makes it
  easy to adjust the vacuum parameters by hand.
\end{itemize}

\subsection{Large N expansion}
\hspace*{\parindent}

Now we use large N expansion to analyze the
vacuum state of supergravity theories.
The extension from the previous analysis (analysis for global
models) is straightforward.

\subsubsection{Introduction}
\hspace*{\parindent}

In the previous section we have analyzed the large N expansion
and gap-equations for global supersymmetric models.
The key idea of the analysis was to consider an intermediate scale 
effective Lagrangian which couples to R-axion superfield.

In this section, we extend the analysis of global
supersymmetric models with R-symmetry to its local version
with Weyl symmetry.
This Weyl symmetry is always present when one consider the superstring 
inspired models and this symmetry needs a compensator superfield.
This compensator
plays almost the same role as R-symmetry compensator 
in the global models. 

\subsubsection{Gaugino condensation in supergravity}
\hspace*{\parindent}  
In the standard superfield formalism of the locally supersymmetric action,
we have the Lagrangian:
\begin{eqnarray}
  \label{lag}
  S&=&\frac{-3}{\kappa^{2}}\int d^{8}z E exp\left(-\frac{1}{3}
  \kappa^{2} K_{0} \right)\nonumber\\
  &&+\int d^{8}z{\cal E}\left[W_{0}+\frac{1}{4}f_{0}{\cal WW}\right]
    +h.c.
\end{eqnarray}
Here we set $\kappa^{2}=8\pi/M_{p}^{2}$.
In the usual formalism of minimal supergravity, the Weyl rescaling is done 
in terms of component fields.
However, in order to understand the anomalous quantum corrections 
to the classical action, we need a manifest supersymmetric formalism,
in which the Weyl rescaling is also supersymmetric.
It is easy to see that the classical action(\ref{lag}) itself is not
super-Weyl invariant.
However,  the super-Weyl invariance can be  recovered
with the help of a chiral superfield $\varphi$(Weyl compensator).

For the classical action (\ref{lag}), the K\"ahler function $K_{0}$, the
superpotential $W_{0}$ and the gauge coupling $f_{0}$ are modified
in the following way\cite{kap}:
\begin{eqnarray}
  \label{modify}
  K_{0} &\rightarrow& K=K_{0}-6\kappa^{-2}{\sf Re} log\varphi \nonumber\\
  W_{0} &\rightarrow& W=\varphi^{3}W_{0}\nonumber\\
  f_{0} &\rightarrow& f=f_{0}+\xi log\varphi
\end{eqnarray}
$\xi$ is the constant which is chosen to cancel the super-Weyl anomaly.
The super-Weyl transformation contains an R-symmetry in its
imaginary part, so we can think that this is a natural extension
of \cite{ross} in which
 a compensator for the R-symmetry played a crucial role
in obtaining the gap equation.

\begin{figure}[hp]
  \epsfxsize=13cm

  \centerline{\epsfbox{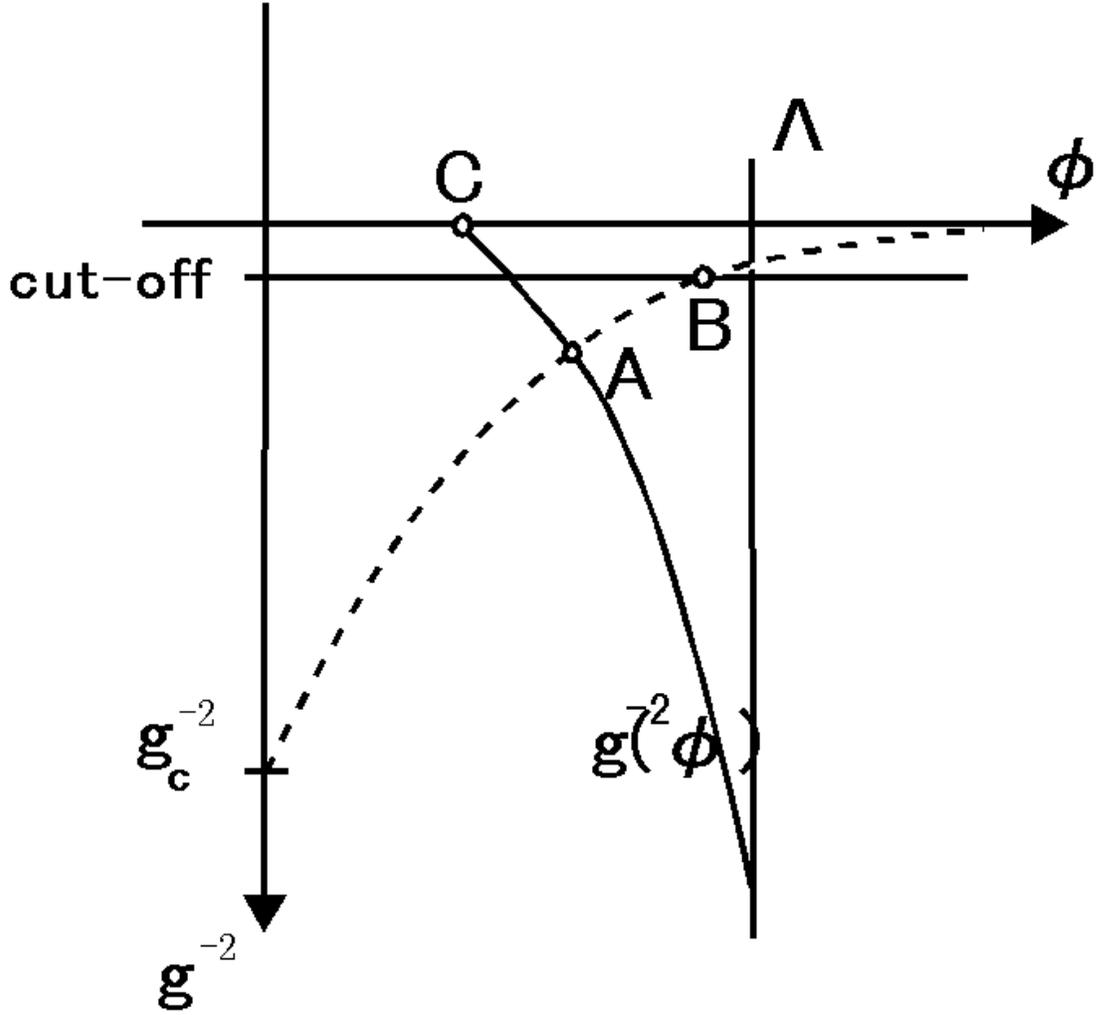}}

 \caption
 {The stationary point of the scalar potential(dotted line) and
   the relation between $g$ and $\varphi$ (solid line) are
   plotted in the same graph.
   The only ``exact'' solution lies at A.
   If we include a cut-off for the four-fermi interaction coefficient
   by hand, we will find a vacuum at B(see Appendix).
   C is the strong coupling point, where $Ref$ vanishes and the 
   gauge coupling $g$ goes to infinity.
   In many cases, we use C for the solution for the non-trivial
   gaugino condensation(The relation between  effective potential
   analysis and the condition $f=g^{-2}=0$ is well discussed in
   ref.[34]).}
 \label{fig:sugra}
\end{figure}

Let us examine the simplest case.
We include some scale factor $\Lambda$  and 
set the form of $W_{0}$ and $f_{0}$ as:
\begin{eqnarray}
  \label{fix}
  W_{0}&=& \lambda \Lambda^{3}\nonumber\\
  Ref_{0}&=& \frac{1}{g^{2}_{0}}
\end{eqnarray}
and rescale the field $\varphi$ as:
\begin{equation}
  \tilde{\varphi}=\Lambda \varphi
\end{equation}
Finally we have:
\begin{eqnarray}
  \label{fin}
  K&=&K_{0}-6\kappa^{-2}{\sf Re}
  log\left(\frac{\tilde{\varphi}}{\Lambda}\right)
  \nonumber\\
  W&=&\lambda {\tilde{\varphi}}^{3}\nonumber\\
  f&=&\frac{1}{g^{2}_{0}}+\xi log\left(\frac{\tilde{\varphi}}{\Lambda}\right)
\end{eqnarray}
The sign of $\lambda$ should respect the condition $\lambda\xi>0$
so that the gap-equation can develop non-trivial solution.
We can obtain a constraint equation from the equation of motion 
for the auxiliary component  $h_{\varphi}$ of the super-Weyl
compensator.
The relevant part of the Lagrangian is:
\begin{equation}
  e^{-1}L_{AUX}=\left[\frac{\partial W}{\partial \varphi}
  -\frac{1}{4}\frac{\partial f}{\partial \varphi}\lambda\lambda
  \right]h_{\varphi}
\end{equation}
From the equation of motion, constraint is now written as:
\begin{eqnarray}
  \frac{\partial W}{\partial \varphi}=\frac{1}{4}
  \frac{\partial f}{\partial \varphi}\lambda\lambda
\end{eqnarray}
for the rescaled field, this can be written as:
\begin{equation}
  \label{aux0}
  \lambda \tilde{\varphi}^{3}-\frac{\xi}{12}
  g^{2}\lambda^{\alpha}\lambda_{\alpha}=0
\end{equation}
The tree level scalar potential  for this minimal model is:
\begin{eqnarray}
  \label{pot1}
  V_{0}&=&-3\kappa^{2}|W|^{2}\nonumber\\
  &=&-3\kappa^{2}\lambda^{2}|\tilde{\varphi}^{3}|^{2}
\end{eqnarray}
The equation of motion for the auxiliary field(\ref{aux0})
 suggests that eq.(\ref{pot1}) can be interpreted as a four-fermion
interaction of  the gaugino:
\begin{equation}
  \label{4f}
  -\frac{1}{48}\kappa^{2}g^{4}
  \xi^{2}|\lambda^{\alpha}\lambda_{\alpha}|^{2}
\end{equation}
This four-fermion interaction becomes strong as $\frac{1}{g^{2}}={\sf Re} f$ 
reaches 0.
The strong coupling point is:
\begin{equation}
  \label{strong}
  \tilde{\varphi}_{s}=\Lambda e^{-\frac{1}{g^{2}_{0}\xi}}
\end{equation}
In the effective Lagrangian analysis, one usually assumes that  this point
is the true vacuum.
We show the vacuum state of three types of analysis(effective Lagrangian
analysis, G.G.Ross type and large N expansion) in Fig.(\ref{fig:sugra}).

\begin{figure}[hp]
  \epsfxsize=13cm

  \centerline{\epsfbox{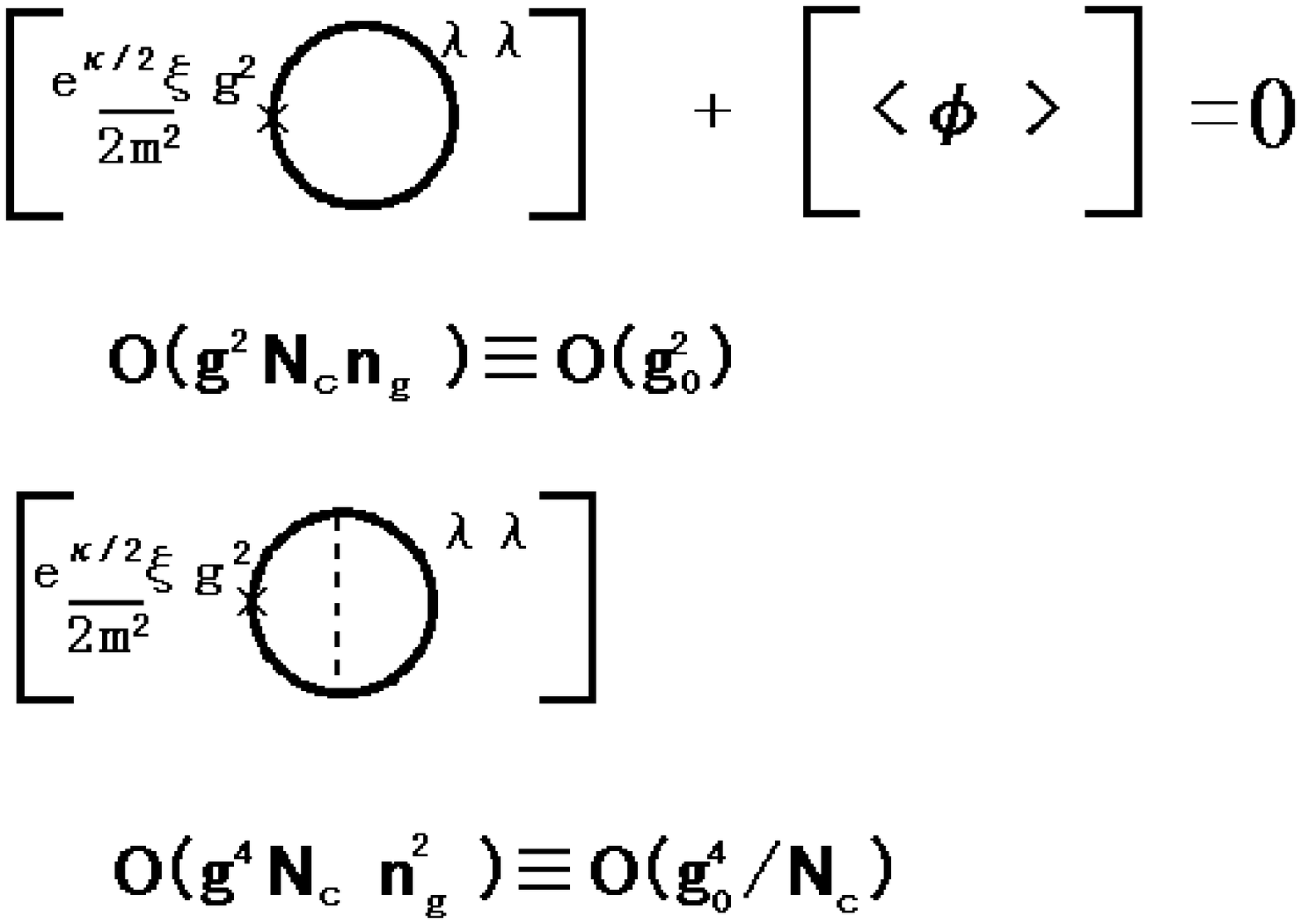}}

 \caption{The gap equation (4.23) is written in a diagrammatic form.
One of the typical higher correction is shown in the second diagram.}

 \label{fig:largeN_sugra}

\end{figure}

By using the tadpole method we can obtain a gap equation:
\begin{eqnarray}
  \label{g}
  \lambda \tilde{\varphi}^{3}&\times&
  \left(1-4G^{2}
    \int \frac{d^{4}p}{(2\pi)^{4}}
    \frac{1}{p^{2}+m_{\lambda}^{2}}\right)=0\nonumber\\
  &&\left\{
    \begin{array}{l}
      G^{2}=\frac{\xi^{2}\kappa^{2}g^{4}n_{g}}{12}\\
      m_{\lambda}^{2}=\frac{\kappa^{4}\xi^{2}g^{4}\lambda^{2}
        |\tilde{\varphi}^{3}|^{2}}{4}
    \end{array}
  \right.
\end{eqnarray}
where $\xi$ is determined by the anomaly constraint and $n_{g}$ is
the dimension of the gauge group($n_{g}=N_{c}^{2}-1$). 
Here, it is proportional to $N_{c}$.
This equation is a good approximation when we 
take $N_{c}\rightarrow\infty$ limit. 
(The leading contribution is enhanced by an extra $N_{c}$ factor.
See fig.(\ref{fig:largeN_sugra})

The solution for the gap equation(\ref{g}) is plotted in Fig.(\ref{fig:sugra}).
We can see that there is always a solution for non-zero 
gaugino condensation.
Now let us consider the difference between our result and ref.\cite{ross}.
In ref.\cite{ross}, the solution for the gap equation is estimated
after fixing the coupling constant at $g_{c}$ which is introduced
by hand.
It is true that the effective potential is singular at $\tilde{\varphi}_{s}$
(\ref{strong}),
but without introducing the cut-off for the strength of the four 
fermion coupling $|\lambda\lambda|^{2}$,
 we can find a stable solution for (\ref{g})
at finite value.(see Fig (\ref{fig:sugra}))


For a second example, we include the dilaton superfield $S$.
Now $f_{0}$ is not a constant and depends on the field $S$:
\begin{equation}
  f_{0}=S
\end{equation}
And the K\"ahler potential for the dilaton superfield is:
\begin{equation}
  K_{0}=-\kappa^{-2}log(S+\overline{S})
\end{equation}
Here we should include the effect of the 
 dilaton field in the scalar potential.
The tree level scalar potential is:
\begin{equation}
  V_{0}=h_{S}(G^{-1})^{S}_{S}h^{S}-3\kappa^{2}|W|^{2}
\end{equation}
The auxiliary field for $S$ is:
\begin{eqnarray}
  \label{aux}
  h_{S}&=&\kappa^{2}\left[\frac{1}{2} \frac{W}{S+\overline{S}}
  +\frac{1}{4}f_{S}
  \lambda^{\alpha}\lambda^{\alpha}\right]\nonumber\\
  &=&\frac{\kappa^{2}}{4}W\frac{1+12S_{R}\xi^{-1}}{S_{R}}
\end{eqnarray}
Here we set $G=K+ln(\frac{1}{4}|W|^{2})$ and $S_{R}=(S+\overline{S})/2$.
The tree level potential can be given in a simple form
\begin{equation}
  \label{sc}
  V_{0}=\lambda^{2}A|\tilde{\varphi}^{3}|^{2}
\end{equation}
where
\begin{equation}
  A=\frac{1}{16}
  \kappa^{2}\left[\left(1+\frac{12S_{R}}{\xi}\right)^{2}
  -3\right].
\end{equation}
The tree level potential for $\tilde{\varphi}$ has no stable 
supersymmetry breaking solution.
This is consistent with an observation that gaugino condensation
as usually parameterized does not occur in models with a single
 gauge group in the hidden sector.
However, we have argued that it is essential to go beyond
tree level to include non-perturbative effects in the effective potential
which may allow non-trivial minimum even in the simplest case
of a single hidden sector gauge group.
This non-perturbative sum is readily obtained by computing
the one-loop correction to $V$, or directly from the gap-equation.
In this case, the gap equation is given by the same constraint equation
(\ref{aux0}), but gaugino mass term is modified.
\begin{eqnarray}
  \label{gg}
  \lambda \tilde{\varphi}^{3}&\times&
  \left(1-4G^{2}
    \int \frac{d^{4}p}{(2\pi)^{4}}
    \frac{1}{p^{2}+m_{\lambda}^{2}}\right)=0\nonumber\\
  &&\left\{
    \begin{array}{l}
      G^{2}=\frac{\lambda g^{4}\xi^{2}g^{4}n_{g}A}{24}\\
      m_{\lambda}^{2}=\frac{A^{2}\xi^{2}g^{4}
        \lambda^{2}g^{2}|\tilde{\varphi}^{3}|^{2}}{36}\\
      n_{g}=N_{c}^{2}-1
    \end{array}
  \right.
\end{eqnarray}
This equation is, however, too complicated to find a
solution.
We thus forced to employ some simplification of
this analysis, for example, fix the four-fermi
interaction coefficient\cite{ross} or to approximate
the solution at which $g^{2}$ goes to infinity, i.e.
$f(S)=0$\cite{leff}.
Here we only refer to the paper\cite{ross} in which
the stability of the potential is well discussed after
fixing the four-fermi interaction coefficient.
In \cite{ross}, duality invariant Lagrangian is also 
considered and it is found that this type of approach
takes into account some additional non-perturbative effects
\cite{add}
and stabilizes the dilaton potential 
with only one gauge group.

To see the basic mechanism involved, let us reconsider eq.(\ref{sc}).
This potential suggests that, once gaugino condensate and
$\varphi$ gets non-zero value, $S$ cannot run away to infinity.
On the other hand, it is obvious from fig.(\ref{fig:sugra})
that eq.(\ref{gg}) always have non-trivial (condensating)
solution for any $S$.
Moreover, in the effective Lagrangian analysis one 
consider that the point C in fig.(\ref{fig:sugra}) is
the solution for non-trivial gaugino condensation.
As $S$ becomes large, C moves toward the origin as the function
of $e^{-S}$ so one tends to think that the potential (\ref{sc})
is destabilized.
On the other hand, if one uses Nambu-Jona-Lasinio type approach\cite{ross},
one is lead to the solution B in fig.(\ref{fig:sugra}) and finds 
that the potential is indeed stable.
However, taking large N limit, we can find that these two solutions
corresponds to different kinds of approximations for the true solution
at A, which is different from both B and C.
Because the solution A is very similar to the solution B 
and does not behave as $e^{-S}$,
 we may think that  the potential is stabilized at the true vacuum A.

\subsection{Summary of section 4}
\hspace*{\parindent}
In this section we have considered the analysis 
of the non-perturbative properties
of supergravity theory which is induced by low energy
dynamics of gaugino condensation.
After a brief review of the previous approaches,
we examined these known results from another point of view.
By using large N expansion, we showed that the effective Lagrangian 
analysis and Nambu-Jona-Lasinio type approach can be
regarded as different kinds of approximations to the
exact solution.

In the large N limit, we can easily understand  why the dilaton 
potential is stabilized in Nambu-Jona-Lasinio type approach but
not stabilized in the effective Lagrangian approach.
We also found a stable and exact solution in the large N limit.
It is important  that we can find a stabilized dilaton potential
if we take into account a certain non-perturbative effect from supergravity.

\newpage

\newpage

\newpage
\section{Conclusions and discussions}
\hspace*{\parindent}
In this paper, we have given a systematic study of the dynamical
properties of supersymmetric theories by using the large N expansion.

First we have analyzed the O(N) non-linear sigma model as
the simplest and explicitly solvable example.
All the parameters are determined exactly and no ambiguities are left.
We have also examined the effects of the supersymmetry breaking 
mass term on the dynamical properties.

For the second example, we have studied supersymmetric QCD
with $N_{f}<N_{c}-1$.
What we have been concerned with was the condensation of gaugino 
which can be viewed as the source of the non-perturbative 
superpotential.
Large N expansion is realized in the limit of $N_{f}\rightarrow\infty$
while $N_{c}-N_{f}$ is fixed.
The results  almost coincide with the previous analyses\cite{rep,ADS}.
We have examined a non-trivial assumption on phase transition
for gaugino condensation that was made in ref.\cite{ADS}
and given alternative proof for it.
We have proved in the large N limit
that no phase transition is allowed for  gaugino condensation in 
 massive SQCD.
However, in massless SQCD, phase transition does occur in the
weak coupling region.
This can change the asymptotic form of the runaway potential.
To be more precise,  the flat directions are only partially lifted
and  does not run away.

For the third example, we have analyzed superstring motivated
supergravity theories  which is the main theme of our paper.
In the large N limit, we found that the analyses derived from
effective (confined) theory\cite{leff} and one from
Nambu-Jona-Lasinio type approach\cite{ross} correspond to
the different kinds of approximations for the exact solution.
(This is explicitly visualized in fig.(\ref{fig:sugra}).)
We have used the constituent fields.
This type of approach
is very important when we think of the phenomenological
implications of supersymmetry breaking mechanism on the
moduli stabilization\cite{add} and the phase transitions
in the early universe\cite{matsuda_cosmo}.
As is discussed in ref.\cite{add}, one should discard  important
part of the non-perturbative effects from supergravity if
he considers only a composite type Lagrangian that should be
obtained by merely turning off the supergravity effects.

\section*{Acknowledgment}
We thank K.Fujikawa  for many helpful discussions.

\appendix
\newpage
\section{Review of the approach by G.G.Ross et al}
\hspace*{\parindent}
In this appendix we review the series of papers written by
G.G.Ross et al\cite{ross}.
The main idea of their analysis comes from an effective low-energy
(still not confined) theory describing the  Goldstone mode 
associated with
the R-symmetry breaking driven by gaugino condensation.
This theory has four-fermi interaction at the tree level
and they examined its implications to gaugino condensation
a la Nambu-Jona-Lasinio type approach.

First, let us construct the Lagrangian.
Demanding that the effective theory given in terms of  the 
auxiliary field $\Phi$ 
generates four-fermion interaction then the form of the $W$ and $f$
are determined.
\begin{eqnarray}
  W&=&m^{2}\Phi\\
  f&=&\xi ln(\Phi/\mu) + S
\end{eqnarray}
where $m$ and $\mu$ are mass parameters, $\xi$ is a dimensionless
constant and  $S$ is a dilaton  superfield. 
Here we take $M_{p}=1$ for simplicity.
The classical equation of motion for auxiliary component of $\Phi$
({$\phi,\chi,h$}) is:
\begin{eqnarray}
  \frac{1}{2}\frac{\partial W}{\partial \phi}&=&\frac{e^{K/2}}{4}
  \frac{\partial f}{\partial \phi}\lambda\lambda
\end{eqnarray}
This gives the relation:
\begin{equation}
  \phi=\frac{e^{K/2}\xi}{2m^{2}}\lambda\lambda
\end{equation}
which means that the field $\phi$ can be an order parameter for
gaugino condensation.
The four-fermion term appears if we consider the tree level 
(supergravity) potential:
\begin{equation}
  V_{0}=3e^{-G}+h_{S}f^{S}(G^{-1})^{S}_{S}
\end{equation}
where $h_{S}$ denotes the F-component of the chiral superfield
$S$ given by:
\begin{equation}
  h_{S}=e^{-G/2}G_{S}+\frac{1}{4}\lambda\lambda-G^{jk}_{S}
  \overline{\chi}_{j}\chi_{k}-\frac{1}{2}\overline{\chi}_{S}
  G_{j}\chi^{j}
\end{equation}
Including all the components, we can write the tree level potential
explicitly:
\begin{eqnarray}
  V_{0}&=&\frac{1}{4}\left(1+\frac{S+\overline{S}}{\xi}\right)^{2}
  e^{-K}|W|^{2}\nonumber\\
  &=&\frac{1}{4}\left(1+\frac{S+\overline{S}}{\xi}\right)^{2}
  e^{-K}m^{4}\phi^{2}
\end{eqnarray}
In terms of gaugino field, we can obtain the four-fermion interaction
term.
\begin{equation}
  L_{4fermi}=\frac{1}{16(Ref)^{2}}\left(1+\frac{S+\overline{S}}{\xi}\right)^{2}
  |\lambda\lambda|^{2}
\end{equation}
where the factor of $Ref$ in the denominator appears
because we have rescaled the gaugino fields appearing in this equation
to have canonical kinetic terms.

The form we have derived depends on the parameters $\xi$, $m$ and $\mu$.
One can determine $\xi$  by demanding that
the low-energy effective Lagrangian is anomaly free under the R-symmetry
transformation.
The mass scale $m$ should be identified with the Planck mass.
Different choices of $\mu$ are possible and we take $\mu=\Lambda^{3}_{GUT}
/m_{Planck}^{2}$.
This choice of $\mu$ is justified in superstring motivated analysis.
A supersymmetry breaking solution to the mass gap $\partial (V_{0}+V_{1})
/\partial \phi=0$ is dynamically favored for a large coupling constant.
The apparent singularity when $Ref$ approaches zero and the potential 
 unbounded from below are not physically reasonable.
On dimensional grounds we introduce a cut-off at the scale
$Ref=\Lambda_{c}$ which corresponds to the effective four-gaugino interaction
$|\lambda\lambda|^{2}/\Lambda_{c}^{2}$.
(See Fig.(\ref{fig:sugra}).)

We can easily  extend this model to include other moduli fields and
string threshold corrections.
Here we give the  explicit form of typical Lagrangian:
\begin{eqnarray}
  K&=&-ln\left(S_{R}+2\sum_{i}k_{a}\delta^{i}_{GS}lnT_{Ri}
  \right)-\sum_{i}lnT_{Ri}\nonumber\\
  W&=&W_{0}+W_{m}\nonumber\\
  f_{0}&=&S+2\sum_{i}(b_{a}^{'i}-k_{a}\delta^{i}_{GS})ln[\eta(T_{i})]^{2}
\end{eqnarray}
where $W_{0}$ is the scalar potential due to the gaugino condensate and
$W_{m}$ is the matter superpotential.

If we demand that the effective theory given in terms of the auxiliary
superfield $\Phi$ generates four-fermi interaction then the form of the
$W$ and $f$ are uniquely determined.
\begin{eqnarray}
  W_{0}&=&m^{2}\Pi_{i}\eta^{-2}(T_{i})\Phi\nonumber\\
  f&=&f_{0}+\xi ln\frac{\Phi}{\mu}
\end{eqnarray}
From the classical equation of motion the scalar component of the
auxiliary superfield $\Phi$ is given in terms of the gaugino
bilinear by:
\begin{equation}
  \phi=\frac{e^{-K/2}\xi}{2m^{2}\Pi_{i}\eta^{-2}_{i}}\lambda\lambda
\end{equation}
For a pure gauge theory in the hidden sector the tree level scalar
potential is now written by:
\begin{eqnarray}
  V_{0}&=&m^{2}_{\frac{3}{2}}B_{0}\nonumber\\
  B_{0}&=&\left(1+\frac{Y}{\xi}\right)^{2}+\sum_{i}\frac{Y}{Y+a_{i}}
  \left(a_{i}-\frac{a_{i}}{\xi}\right)^{2}\frac{T_{Ri}}{4\pi^{2}}
    |G_{2}(T_{i})|^{2}-3
\end{eqnarray}
where $G_{2}$ is the Einstein modular form with weight $1/2$ and
$a_{i}$ is defined as $a_{i}=2(k_{a}\delta^{GS}_{i}-b^{'}_{ai})$.
The gravitino mass is given by:
\begin{equation}
  m_{\frac{3}{2}}=\frac{1}{4}e^{K}\Pi_{i}|\eta(T_{i})|^{-4}|\phi|^{2}
\end{equation}
It is clear that a non-zero gravitino mass is only possible for
non-vanishing VEV of $\phi$.
The one-loop radiative corrections may be calculated using Coleman-Weinberg
one-loop effective potential.
\begin{equation}
  V_{1}=\frac{1}{32\pi^{2}}Str\int d^{2}p p^{2} ln(p^{2}+M^{2})
\end{equation}
After introducing the cut off parameter for $Ref$ and fixing the
form of four-fermion interaction term as 
$|\lambda\lambda|^{2}/\Lambda_{c}^{2}$, the extremum conditions are
solved.

Here we do not describe further  detailed analysis on this phenomenological
applications because our aim in this appendix is to introduce
a main idea of the method developed by G.G.Ross et al.

\end{document}